\documentclass[12pt,preprint]{aastex}
\usepackage{pslatex,color}
\usepackage{natbib}
\usepackage{amsmath,amsfonts,amssymb,amsxtra,url}
\usepackage{ulem}
\usepackage{color}

\usepackage{color}
\definecolor{myred}{rgb}{0.7,0.1,0.1}
\definecolor{myblue}{rgb}{0.2,0.0,0.7}
\definecolor{mybrown}{rgb}{0.5,0.2,0.0}

\newcommand\hide[1]{}
\shorttitle{Predicting a third planet in the Kepler-47 circumbinary system}
\shortauthors{Hinse, T. C. et al.}

\begin{document}

\title{Predicting a third planet in the Kepler-47 circumbinary system}

\author{Tobias C. Hinse}
\affil{Korea Astronomy \& Space Science Institute, Republic of Korea}
\affil{Armagh Observatory, College Hill, BT61 9DG, Northern Ireland, UK}
\email{tchinse@gmail.com}
\author{Nader Haghighipour}
\affil{Institute for Astronomy, University of Hawaii-Manoa, Honolulu, Hawaii, USA}

\author{Veselin B. Kostov}
\affil{Johns Hopkins University, Baltimore, USA}

\author{Krzysztof Go{\'z}dziewski}
\affil{Toru{\'n} Centre for Astronomy of the Nicolai Copernicus University, Grudziadzka 5, Poland}
%
\begin{abstract}
We have studied the possibility that a third circumbinary planet in the Kepler-47 planetary system 
be the source of the single unexplained transiting event reported during the discovery of these planets. 
We applied the MEGNO technique to identify regions in the phase space where a third planet can maintain 
quasi-periodic orbits, and assessed the long-term stability of the three-planet system by integrating the
entire 5 bodies (binary + planets) for 10 Myr. We identified several stable regions between the two known 
planets as well as a region beyond the orbit of Kepler-47c where the orbit of the third planet could be
stable. To constrain the orbit of this planet, we used the measured duration of the unexplained transit 
event ($\sim 4.15$ hours) and compared that with the transit duration of the third planet in an ensemble 
of stable orbits. To remove the degeneracy among the orbits with similar transit durations, we considered 
the planet to be in a circular orbit and calculated its period analytically. The latter places an upper limit
of 424 days on the orbital period of the third planet. Our analysis suggests that if the unexplained transit
event detected during the discovery of the Kepler-47 circumbinary system is due to a planetary object,
this planet will be in a low eccentricity orbit with a semimajor axis smaller than 1.24 AU.
Further constraining of the mass and orbital elements of this planet requires a re-analysis of the
entire currently available data including those obtained post-announcement of the discovery of this system.
We present details of our methodology and discuss the implication of the results.

\end{abstract}
%
%

\section{Introduction}

During the past few years, {\it Kepler} telescope has discovered several planets in circumbinary orbits. All these 
planets have been detected photometrically, exhibiting transit signatures when passing in front of the stars of 
the binary. The first of these circumbinary planets (CBPs) was Kepler-16b discovered by Doyle et al in 2011. 
Since then several more {\it Kepler} CBPs have been discovered, namely Kepler-38b \citep{Orosz2012a}, Kepler-34b and Kepler-35b 
\citep{Welsh2012}, Kepler-47b\&c \citep{Orosz2012b}, Kepler-64b \citep{Kostov2013, Schwamb2013}, 
Kepler-413b (KIC 12351927b) \citep{Kostov2014}, and KIC 9632895b \citep{Welsh2014}. 

Among the currently known {\it Kepler} circumbinary planetary systems, Kepler-47 presents an interesting
case. The fact that this system harbors two planets is a strong indication that similar to planet formation around single stars, 
circumbinary planets can also form in multiples. The latter implies that more planets may exist in any of the currently
known circumbinary systems. In the Kepler-47 system, the existence of a third planet was speculated in its 2012 discovery paper  
as a way to account for an unexplained feature observed in the light curve of this binary \citep{Orosz2012b}.
As reported by these authors, a single, 0.2\% deep transit event had been detected 
which could not be explained by the two known transiting planets. 

In this paper, we plan to test the above-mentioned hypothesis. Our approach is to study the dynamics of the three-planet system,
and use long-term stability to identify the viable regions where the orbit of the third planet would be stable. 
Using dynamical stability to predict additional planets has been presented in several other studies
\citep{BarnesRaymond2004,RaymondBarnes2005,FangMargot2012}. In cases where stable regions are identified, we will use transit 
timing and transit duration variations to constrain the orbit of the third planet. 

The dynamics and orbital stability of planets in circumbinary orbits
have been the subject of studies for close to three decades. \citet{Dvorak1986}, \cite{RablDvorak1988}, and
\citet{HolmanWiegert1999} carried out long-term orbital integrations of test-particles aiming
at exploring a large area of the system's parameter space. In particular, for P-type orbits, the authors established 
stability criteria for a planet's semimajor axis as a function of the binary orbital parameters and mass-ratio. 
\cite{Musielak2005} and \cite{Eberle2008} also studied the stability of planetary orbits in P-type systems and presented
criteria for stable, marginally stable, and unstable circular planetary orbits.
Recent analytic analysis of the dynamics of circumbinary planets has also been presented by \citet{DoolinBlundell2011},
and \citet{LeungLee2013}.

This paper is structured as follows. In section 2, we briefly review the Kepler-47 system as described by \cite{Orosz2012b}. 
In section 3, we describe our numerical techniques and in section 4, we present the results of our stability analysis 
using the chaos indicator MEGNO (Mean Exponential Growth factor of Nearby Orbits). In section 5, we calculate the transit 
durations of Kepler-47b and the candidate third planet, and compare them with their measured value as reported by 
\cite{Orosz2012b} to constrain the orbit of the third planet. Finally, in section 6, we conclude this study by presenting 
a summary and discussing the implications of the results.

\section{The Kepler-47 system}

Kepler-47 is a single-lined spectroscopic binary with a $1.043~M_{\odot}$ primary and a secondary with a mass
of $0.362~M_{\odot}$. The period of this binary is 7.5 days. In 2012, Orosz et al. announced the detection of two planets in 
circumbinary orbits around this system. These authors analyzed long- and short-cadence photometric data from {\it Kepler}
space telescope, spanning 1050.5 days from Quarter 1 to 12, and identified eighteen transit events of the inner planet 
(Kepler-47b) and three for the outer planet (Kepler-47c).
Table \ref{planetparams} shows the published (osculating) orbital parameters of Kepler-47 and its two planets. 
We note that because of observational degeneracies, not all orbital parameters can be determined from the 
photodynamical\footnote{https://github.com/dfm/photodynam} model as described in \cite{Orosz2012b}. We also note that 
all orbital elements are in the (geometric) Jacobian coordinate system.

The inner planet, Kepler-47b, with a period of 50 days, is the smaller of the two with a radius of  
$\sim 3$ Earth-radii. \cite{Orosz2012b} estimated that the mass of this planet is 7-10 Earth-masses. Due to the non-detection of ETVs, 
an upper limit of 2 Jupiter-masses can firmly be established for this object. The outer planet, Kepler-47c, has an orbital period 
of $\sim 303$ days with a $\sim 4.6$ Earth-radii. The authors estimated a plausible mass in the range 16-23 Earth-masses. 
The upper limit for the mass of this planet was determined to be 28 Jupiter-masses. 

In their analysis of the light curve of Kepler-47 system, \cite{Orosz2012b} detected a single transit event
that could not be explained by the transits of the two planets.  With a formal significance of $10.5\sigma$, this transit event 
occurred at BJD $2,455,977.363 \pm 0.004$, approximately 12 hours after the last transit of Kepler-47b in the Q12 data set. 
The duration of this transit was observed to be $\sim 4.15$ hours. \cite{Orosz2012b} suggested that if 
this transit event is due to a third planet, given its depth of 0.2\%, the planet must have a radius $\simeq 4.5$ Earth-radii.

The rest of this paper is devoted to examining this hypothesis using dynamical considerations. Integrating
the five-body system of the binary and three planets, we will determine the ranges of the parameters for which the orbit of
a hypothetical third planet will be stable, and using the properties of the above-mentioned single transit event, we will
identify the most probable regions around the binary where the orbit of this planet may exist.

\section{Methodology and Numerical Techniques}

We note that results presented in this work have been obtained from direct numerical integration of the equations 
of motion using the initial conditions shown in Table \ref{planetparams}. The orbit of the binary system is 
fully resolved and the three planets are treated as massless as well as massive objects. 

We adopted two different algorithms for solving equations of motion: The IDL implementation of the Livermore 
Solver (LSODE) which is an adaptive algorithm with a step-size control, and an accurate extrapolation method 
implemented as the Gragg-Bulirsh-Stoer (GBS) algorithm (the ODEX code) \citep{Hairer1993}. The latter is frequently 
used in celestial mechanics and orbit calculations 
\cite[see][and references therein]{GozdziewskiEtAl2012,Gozdziewski2013,GozMig2014}. {\bf Both algorithms use a 
relative and absolute error tolerance parameter to control the integration accuracy. We set these parameters to one part 
in $10^{15}$. When integrating several test orbits of the five-body problem, we obtained identical results using 
both algorithms.

We would like to note that, when transforming orbital elements, we use Jacobi-like coordinates with a mass-parameter 
$\mu = k^2(M_{1}+M_{2}+m_{i})$ for each planet. Here $m_i$ is the mass of the $i$th planet, $M_1$ and $M_2$ represent
the masses of the binary stars, and $k$ is the Gauss gravitational constant. The transformed orbital elements are then 
given relative to the center of mass of the binary system. This approach differs from the usual definition of Jacobi 
elements where the position and velocity of a planet are given relative to the center of mass of all remaining massive 
bodies within its orbit.}

Our stability analysis employs the well-established fast chaos indicator MEGNO
technique \citep{CS1999,CS2000,CGS2003} which enables us to explore the phase 
space topology of the system. The MEGNO technique has found widespread 
applications within dynamical astronomy \citep{Goz2001,Goz2003,Goz2008,Hinse2010,Kostov2013} and is closely related 
to the Fast Lyapunov Indicator (FLI) \citep{Mestre2011}. In brief, MEGNO, shown by $\langle Y\rangle$ here and
throughout the paper, has the following properties. For initial conditions resulting 
in quasi-periodic orbits, $\langle Y\rangle \rightarrow 2.0$ for $t \rightarrow \infty$. 
For chaotic orbits, $\langle Y\rangle \simeq {2\lambda}/{t}$ for $t \rightarrow \infty$ where $\lambda$ is the 
Maximum Lyapunov Exponent (MLE). In the simulations presented in this paper, 
we chose to stop a given integration when $\langle Y\rangle > 5$. Orbits with 
quasi-periodic time evolution usually assume values of $|\langle Y\rangle - 2| < 0.001$ at the end of the numerical 
integration. We used the MEGNO implementation within the MECHANIC package \citep{Slonina2015},
and considered the integration time for each orbit to be $4.7 \times 10^{4}$ binary periods.

\section{Orbital stability of the third planet}

We begin our stability analysis by adding a hypothetical third planet (hereafter shown by letter d) to the 
Kepler-47 system. Our goal is to identify regions of the parameter space where the third planet can have a 
long-term stable orbit. Throughout our analysis, we start planets Kepler-47b and c at their published 
osculating elements (Table \ref{planetparams}) and set their masses equal to 10 and 23 Earth-masses, respectively. 
The initial orbital orientation of the third planet is taken to be co-aligned with Kepler-47c.  

\begin{figure}
\center
\includegraphics[width=1.0\textwidth]{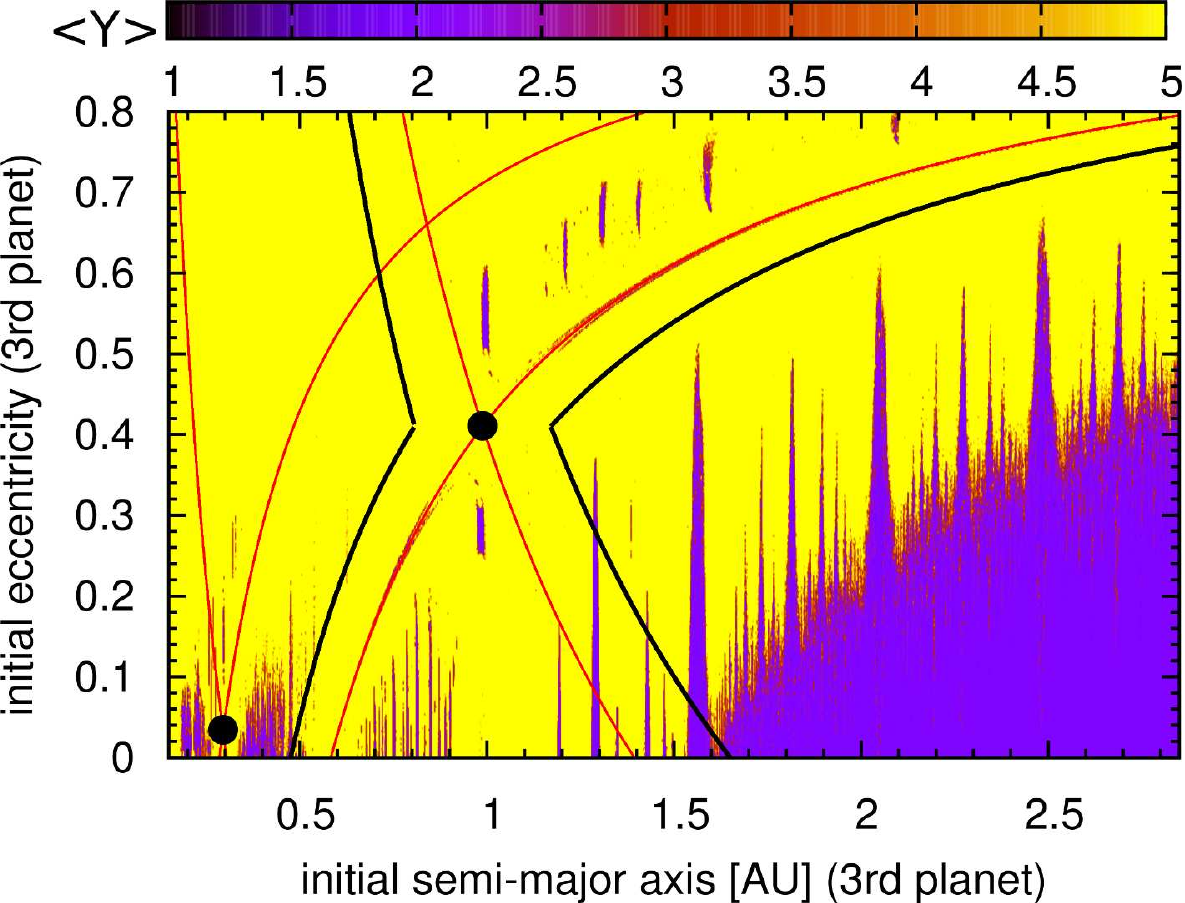}
\caption{Dynamical MEGNO map for the third planet of the Kepler 47 system. The two known planets, Kepler-47b and c 
are shown as black circles. The vertical and horizontal axes correspond to the initial values of the eccentricity and
semimajor axis of the third planet. Yellow color denotes chaotic dynamics and blue indicates quasi-periodic orbits. 
The general quasi-periodic region for a third planet starts from around 1.6 AU. Several quasi-periodic areas for a 
third planet are also embedded in the general chaotic regions between the two known planets. 
Red and black contour lines corresponds to collision (red) and semi-empirical stability criterion (black) 
\citep{Giuppone2013}. See text for more details.}
\label{K47_Map017}
\end{figure}

We first consider the third planet to be massless. Figure \ref{K47_Map017} shows the values of MEGNO for different 
initial values of the semimajor axis and eccentricity of the third planet in an $(a,e)$ space. The initial osculating 
elements for the two known planets are shown by black dots. The color-coding in this and subsequent MEGNO maps
represents the value of MEGNO, $\langle Y\rangle$. Higher values (lighter colors) correspond to higher degree of
chaos and therefore, higher chance of instability. As shown here, there are {\bf three} regions where a third planet can maintain 
a quasi-periodic orbit: {\bf i) a region in the vicinity of Kepler-47b}, ii) a 
region between the two known planets, and iii) the region exterior to the orbit of Kepler-47c (but interior to the general 
quasi-periodic area shown in dark colors).  

\begin{figure}




\center{
\includegraphics[width=0.63\textwidth]{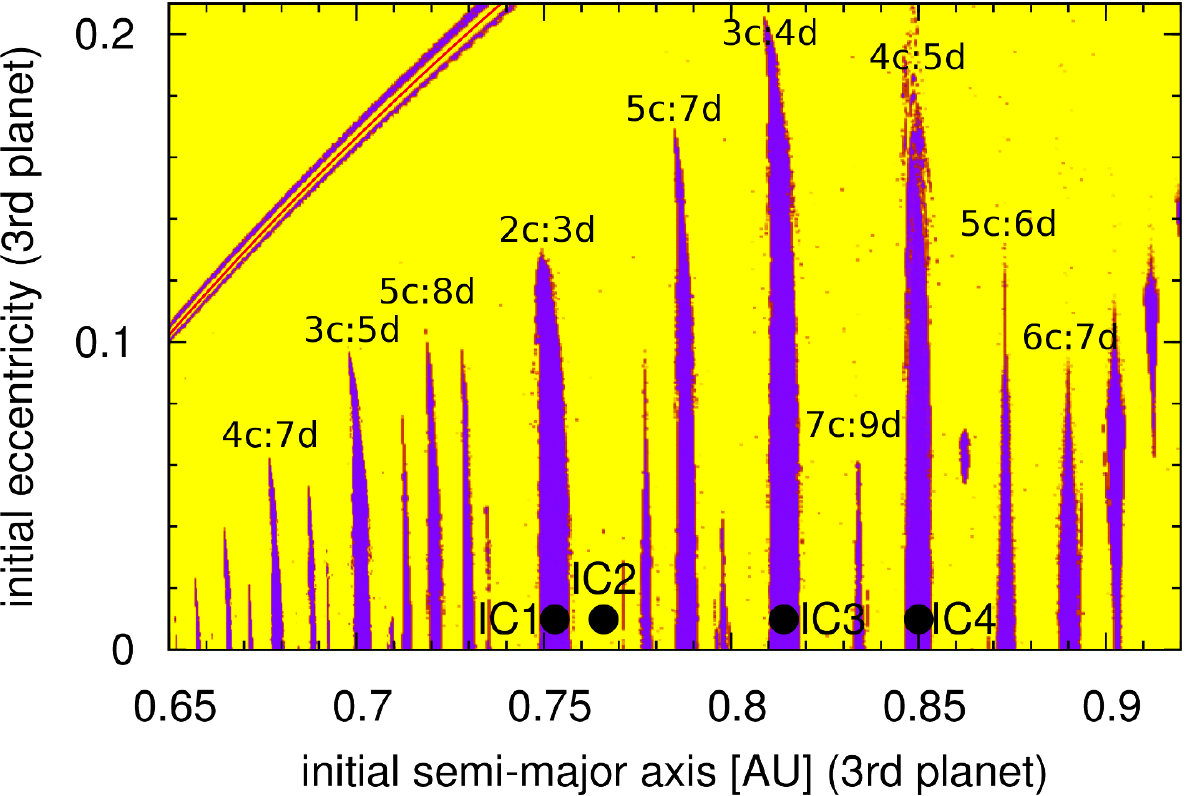}
\vskip 15pt
\includegraphics[width=0.63\textwidth]{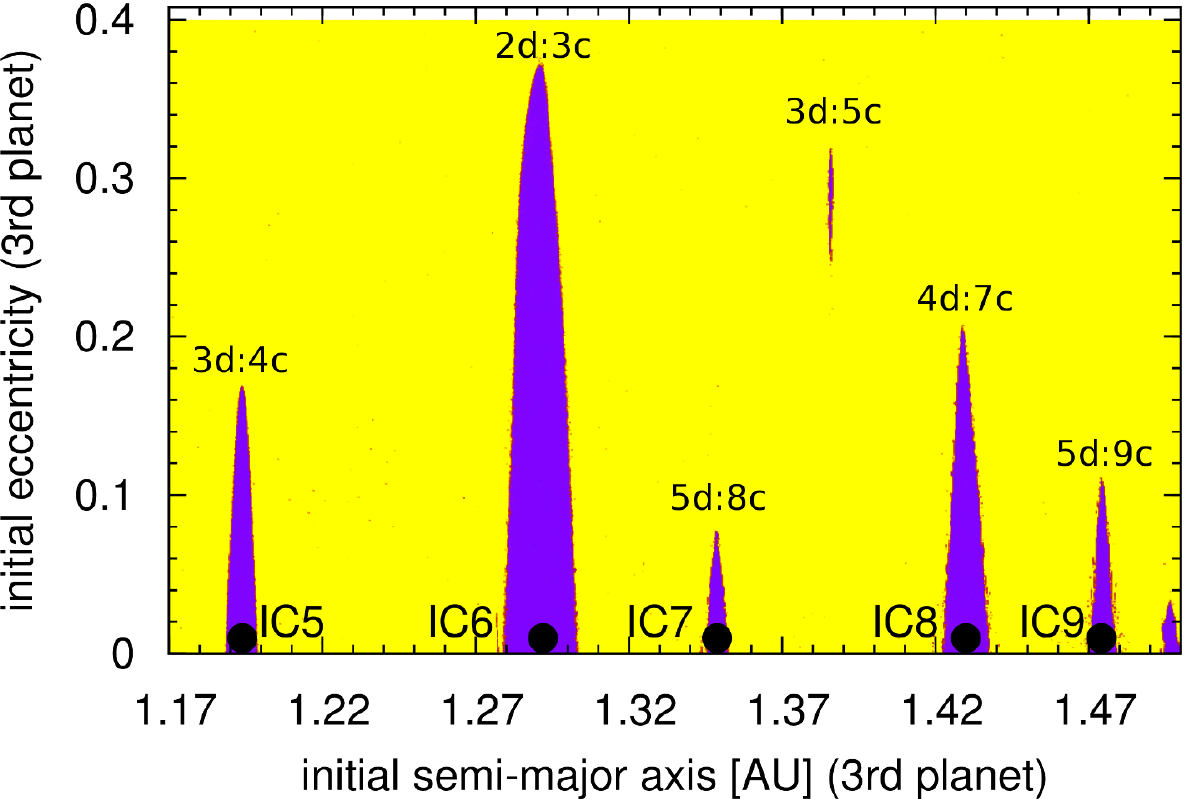}
\vskip 15pt
\includegraphics[width=0.635\textwidth]{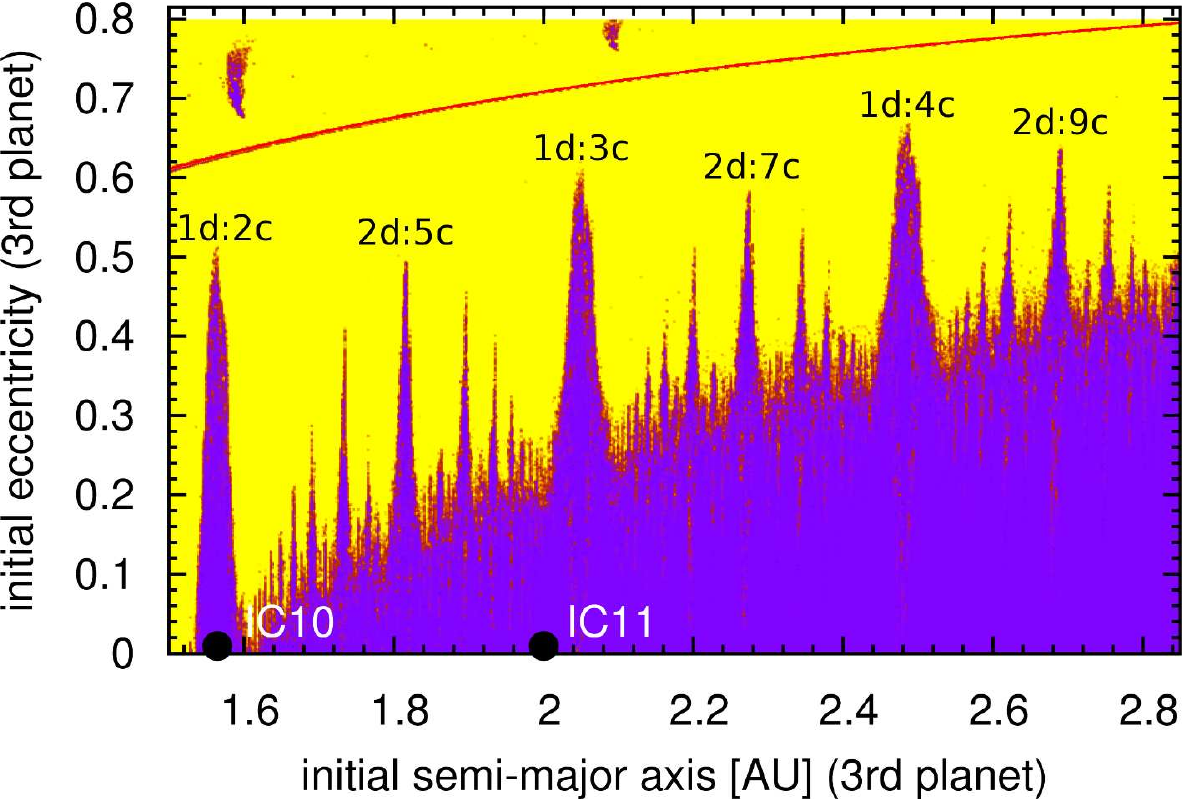}
}
\caption{Same as figure ~\ref{K47_Map017} but zooming into three specific regions. The top panel shows the case where 
the third planet is between Kepler-47b and c. The bottom two panels correspond to the case where the third planet is
outside the orbit of Kepler-47c. The color-coding is the same as that in figure ~\ref{K47_Map017}.}
\label{K47_Map055_Map056}
\end{figure}

{\bf In a multi-planet system, different types of perturbation affect the dynamical evolution of the system on different
timescales. In general, there is a timescale associated with the effect of short-term MMRs, a timescale due to secular resonances 
(slow variation of orbital elements), and a longer evolutionary timescale due to the tidal effects, mass-loss and other weak 
perturbations. When using a fast stability indicators such as MEGNO, because the properties of a given solution in the 
phase-space must be determined on a small length of the orbit, the stability of a particular orbital configuration must 
be presented in the context of these timescales. In a compact configuration, for instance when the third planet is between 
Kepler-47b and c, the dynamics of the system is mainly affected by the short-term 2-body and 3-body MMRs. Similar numerical 
experiments such as those by \citet{GozMig2014} and references therein suggest that in a study as the one presented in this 
paper, when the value of MEGNO converges to $2$ (indicating a quasi-periodic orbit), the stability of the system (the orbit-crossing time) is guaranteed 
for a time 2--3 orders of magnitude longer than the MEGNO integration time-span. In the case of the Kepler-47 system, the 
crossing-time is longer than at least $\sim 5 \times 10^{7}$ binary period. The protecting effect of MMRs imply that 
in some regions such as a number of orbits in the islands of quasi-periodicity at $a_d<1.5$~AU (figure \ref{K47_Map017}),
the entire five-body system will be stable for even longer times (e.g., as long as the lifetime of the system's stars).

It is important to note that when longer-period, multi-body mean-motion or secular resonances exist, 
the convergence of MEGNO for relatively short integration times does not necessarily imply 
stability for the lifetime of the binary. The perturbations from these resonances may result in 
dynamical instability over tens of millions or billions of years \citep[e.g.,][]{LaskarGastineau2009}. That means that either 
the MEGNO integrations must cover a significant fraction of the secular timescales, or direct numerical integrations have to 
be carried out in order to ensure that the system is stable over the time-span of interest.

Despite the above-mentioned shortcoming, MEGNO integrations can still provide accurate and reliable characterization 
of the phase-space with a very small CPU overhead. They are also very useful in obtaining rough stability limits. 
In figure~\ref{K47_Map017}, we have plotted such limits (black curves) around the orbit-crossing curves of the third planet 
with Kepler-47b and c(red curves) following the semi-empirical stability criterion presented by 
\cite{Giuppone2013}. The stability criterion by these authors has been derived from the Wisdom's stability criterion 
for the restricted three-body problem. Unfortunately, this criterion does not seem to determine the limits of stable regions in the Kepler-47
system, properly. For instance, as it will be shown later, the interesting region of $0.5 {\rm (AU)} \leq {a_d} \leq 1.5{\rm (AU)}$ 
is considered to be unstable when using the stability criterion by \cite{Giuppone2013}. However, 
as our MEGNO maps show (see also figure~\ref{K47_Map055_Map056}), this region contains a set of MMRs
where the entire five-body system can be stable. A similar argument applies to the region immediately beyond the orbit 
of Kepler-47c. As indicated by our calculations, a few islands of stability exist in this area that correspond to two-body MMRs 
(e.g., 3d:4c, 2d:3c and 4d:7c), whereas according to the stability criterion by \cite{Giuppone2013}, this region is unstable.}

In figure ~\ref{K47_Map055_Map056}, we show these regions in more detail.
In each panel, we label each quasi-periodic island with its associated two-body mean-motion resonance. Tables \ref{detailsofMMRinterior} and \ref{detailsofMMRexterior} give the values of the inner and outer semimajor axis
of the third planet indicating the width of each resonance valid for the third planet on a circular orbit.

\begin{figure*}
\vbox{
\centerline{
\includegraphics[width=0.5\textwidth]{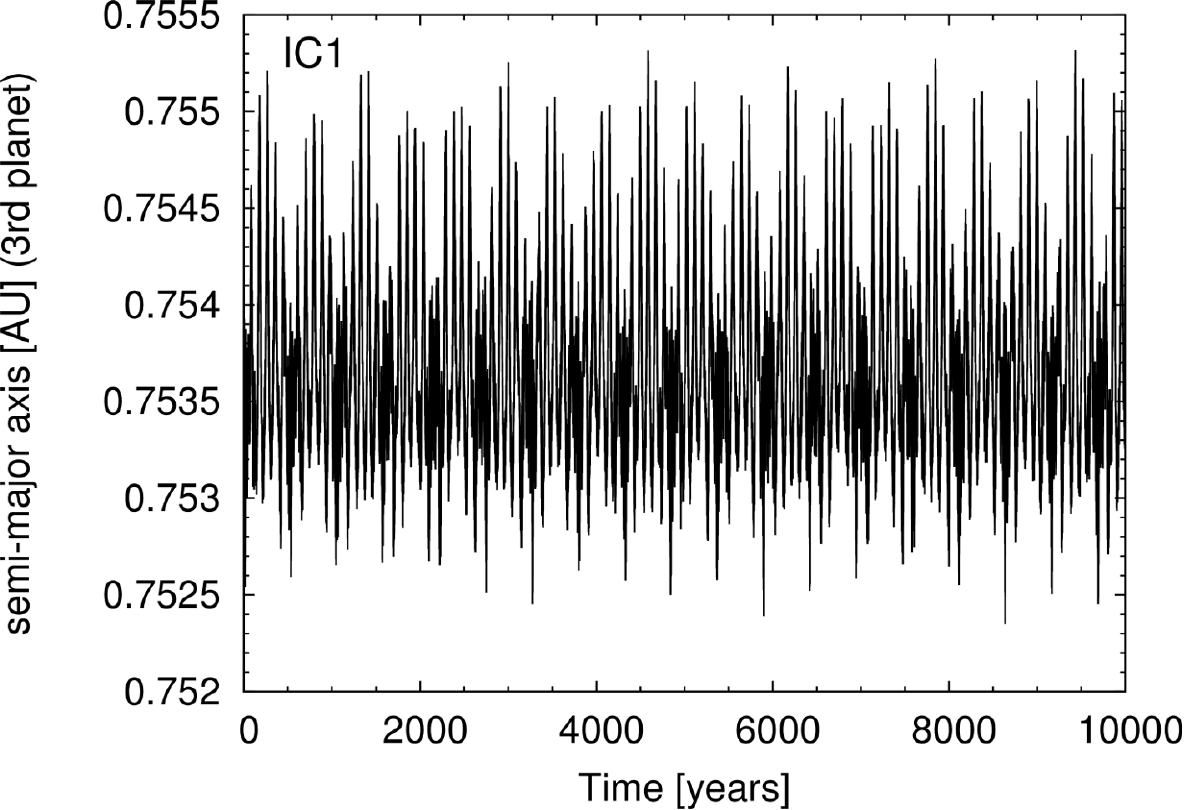}
\includegraphics[width=0.5\textwidth]{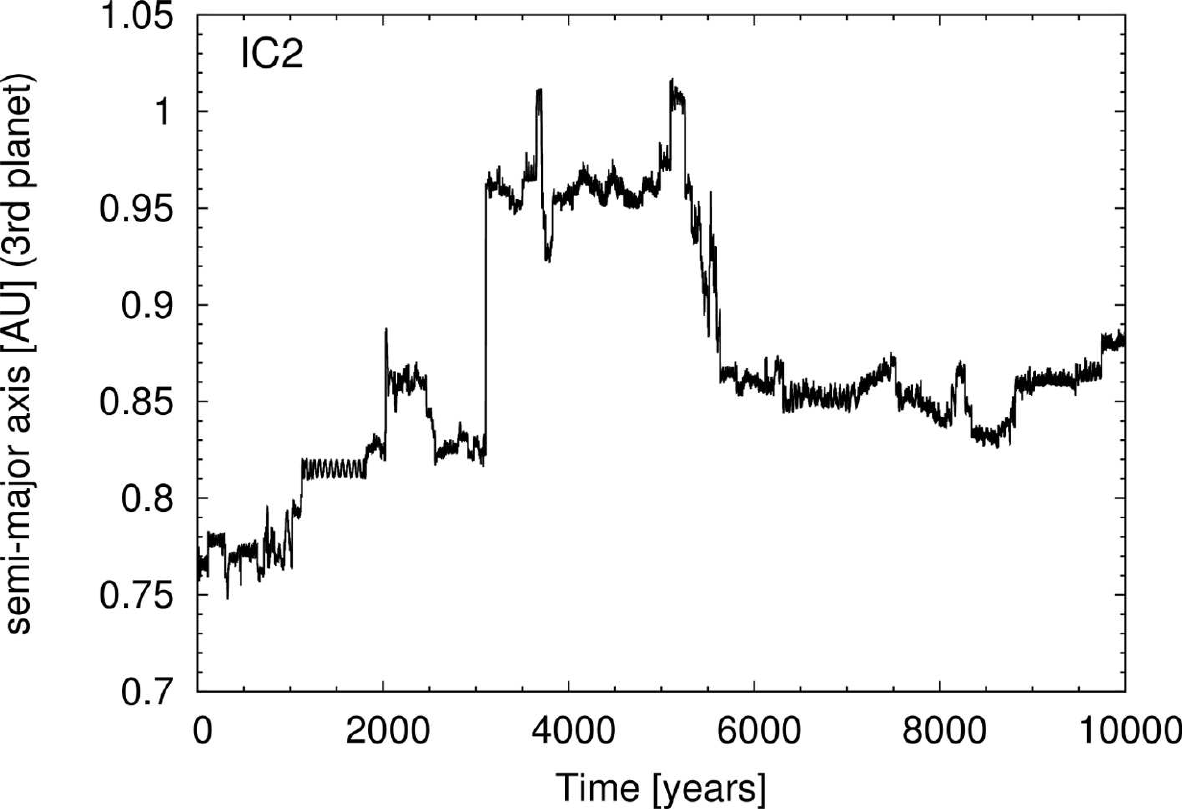}
}
}
\vskip 15pt
\vbox{
\centerline{
\includegraphics[width=0.5\textwidth]{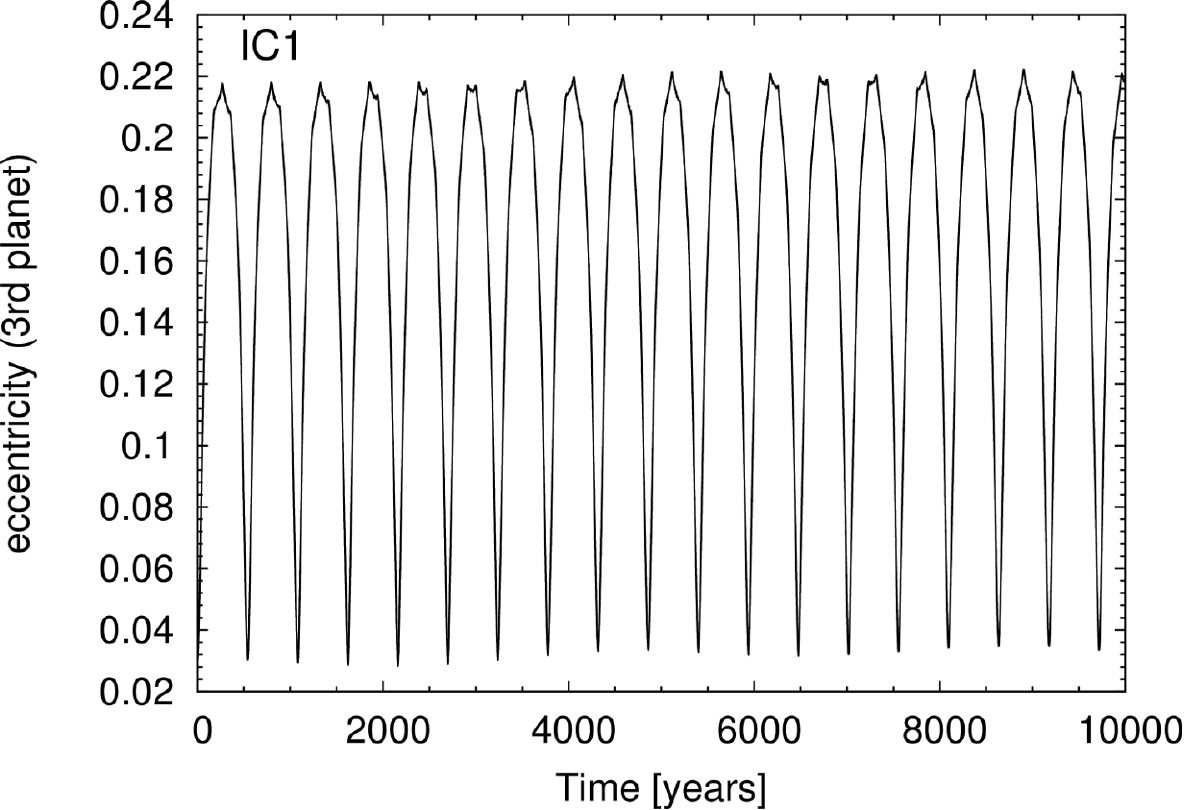}
\includegraphics[width=0.5\textwidth]{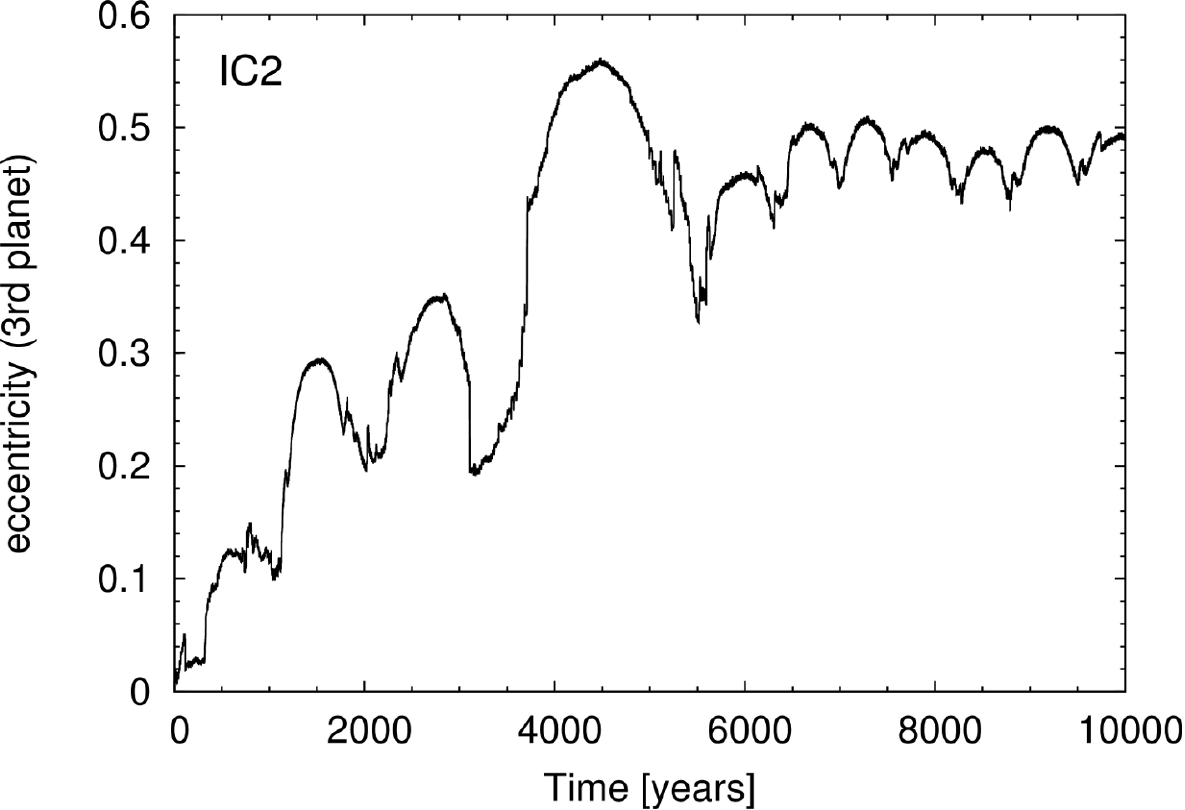}
}
}
\vskip 15pt
\vbox{
\centerline{
\includegraphics[width=0.5\textwidth]{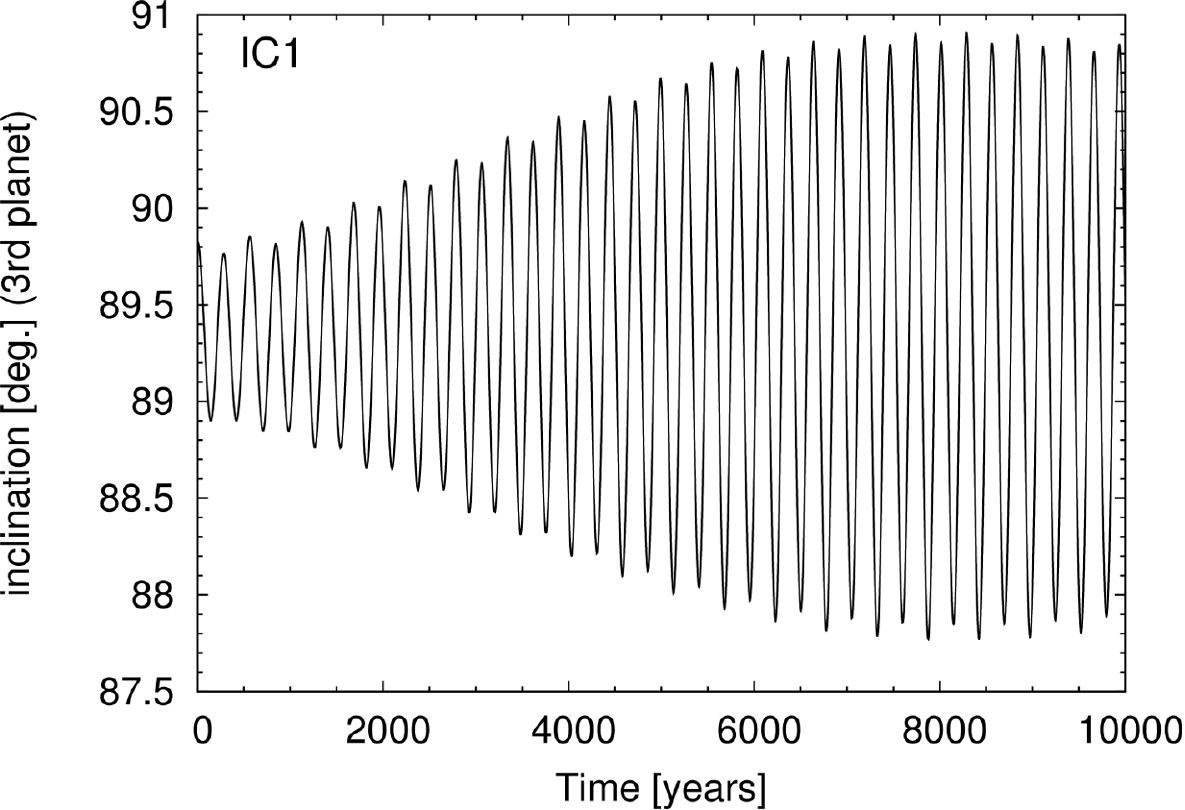}
\includegraphics[width=0.5\textwidth]{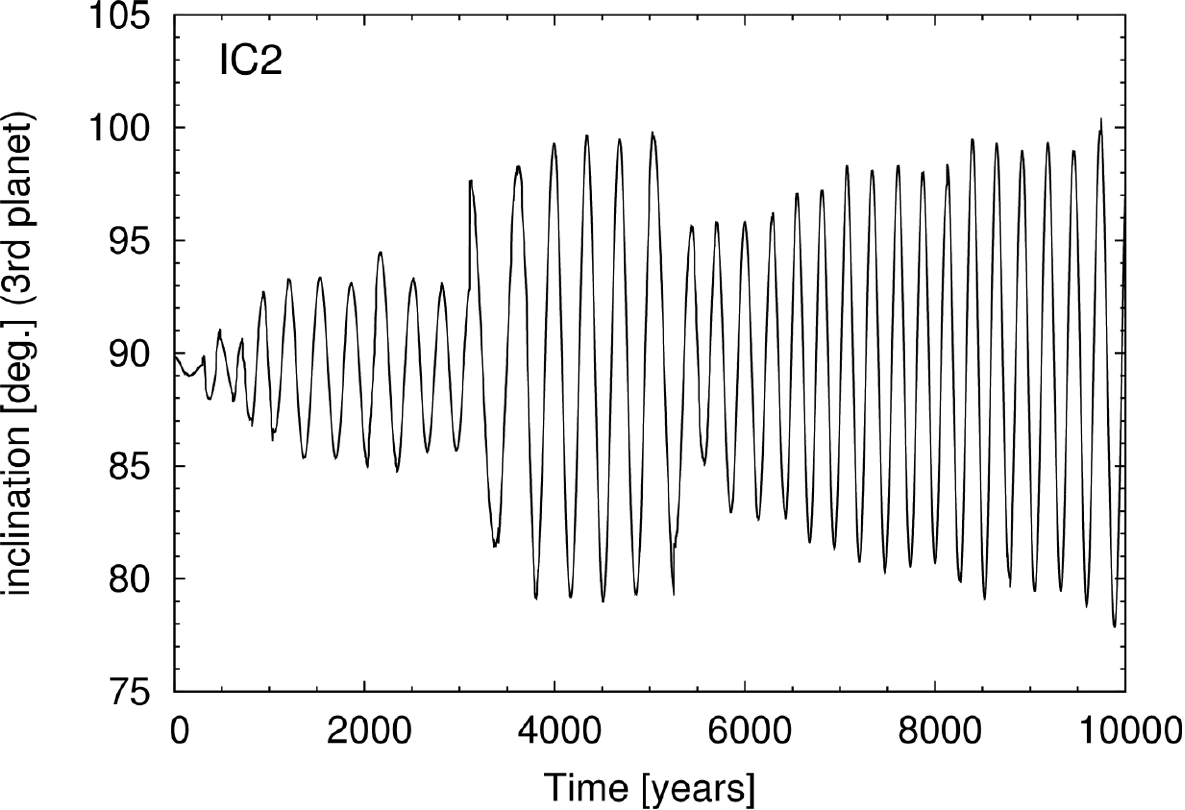}
}
}
\caption{Time evolution of (Jacobian) orbital element of {\bf a third massless companion} considering initial conditions IC1 (left) 
and IC2 (right). Both orbits have initial eccentricity of 0.01. However, for IC2, the initial semimajor axis of the
planet is slightly larger at 0.7657 AU.}
\label{K47_Orbit01_Orbit03}
\end{figure*}

From all the orbits shown in figure \ref{K47_Map055_Map056}, we chose 11 (labelled as IC1 to IC11 standing for Initial 
Conditions 1 to 11) and studied their long-term dynamical evolution by integrating them for $10^7$ years. Results for 
initial conditions IC1 and IC2 are shown in figure ~\ref{K47_Orbit01_Orbit03}. In order to highlight the details of 
the evolution of these orbits, we only show the results of the first 10,000 years of integrations. As shown here, 
both orbits have identical initial eccentricities (0.01). However, their initial semimajor axes 
are different by $\Delta a = 0.0127$ AU. Results indicated that despite such a small difference in initial conditions,
the evolutions of these two orbits are profoundly different. A chaotic characteristics is clearly visible for 
IC2 which exhibits a random walk in semimajor axis and eccentricity. This random walk is an indication of a stochastic 
process over time, slowly destabilizing the orbit by driving the orbital eccentricity to unity. We find the third planet 
to be eventually ejected from the system after some 190,000 yrs. For IC1, the third planet follows a bound stable 
quasi-periodic orbit over at least $10^7$ years and shows no sign of chaotic diffusion. The orbital evolution of 
the two known planets of the system, Kepler-47b and Kepler-47c (both with their assigned masses) show similar dynamical 
behavior with their orbits bound between a minimum and maximum value for their orbital elements.

\subsection{Identification of two-body resonances}

\begin{figure*}
\vbox{
\centerline{
\includegraphics[width=0.5\textwidth]{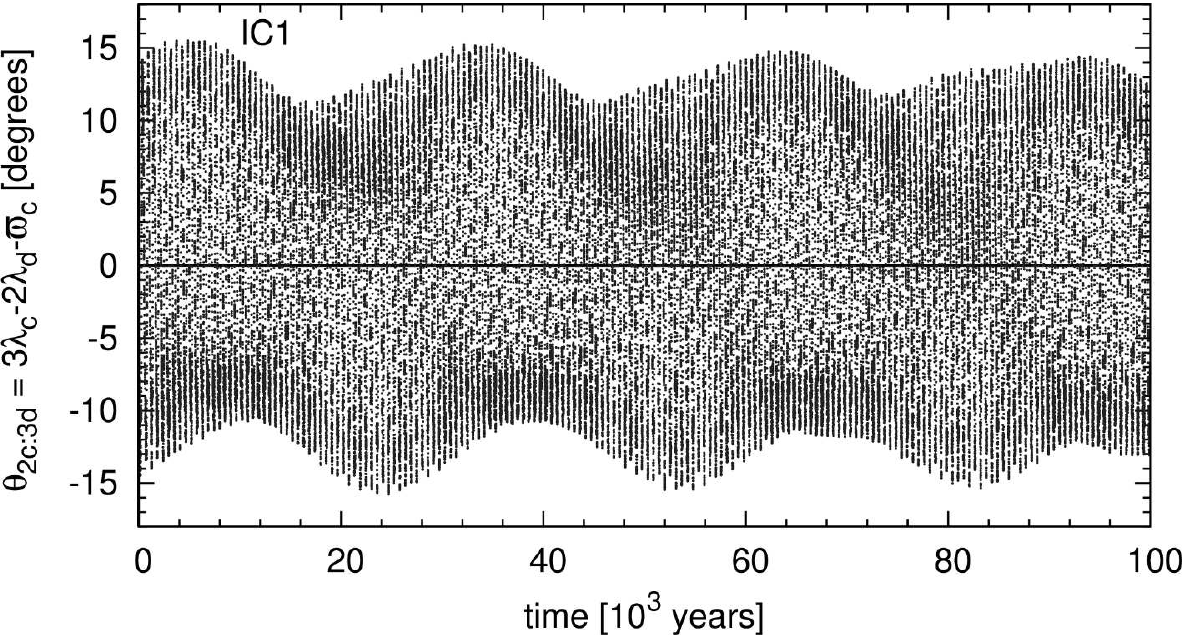}
}
}
\vskip 15pt
\vbox{
\centerline{
\includegraphics[width=0.5\textwidth]{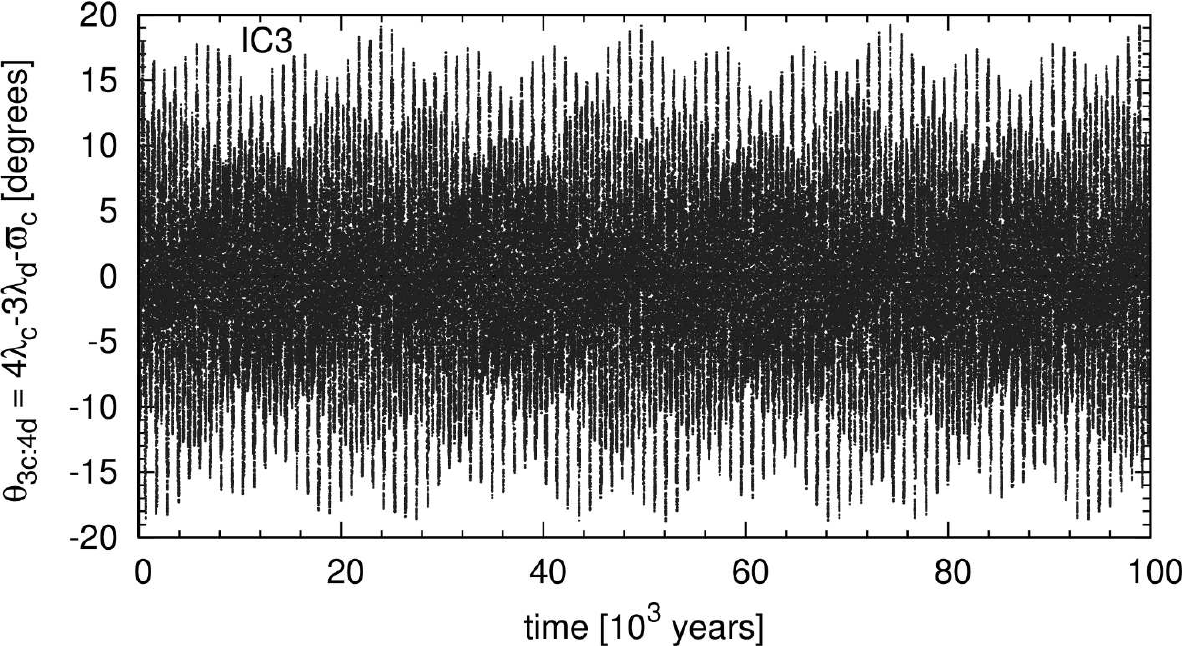}
}
}
\vskip 15pt
\vbox{
\centerline{
\includegraphics[width=0.5\textwidth]{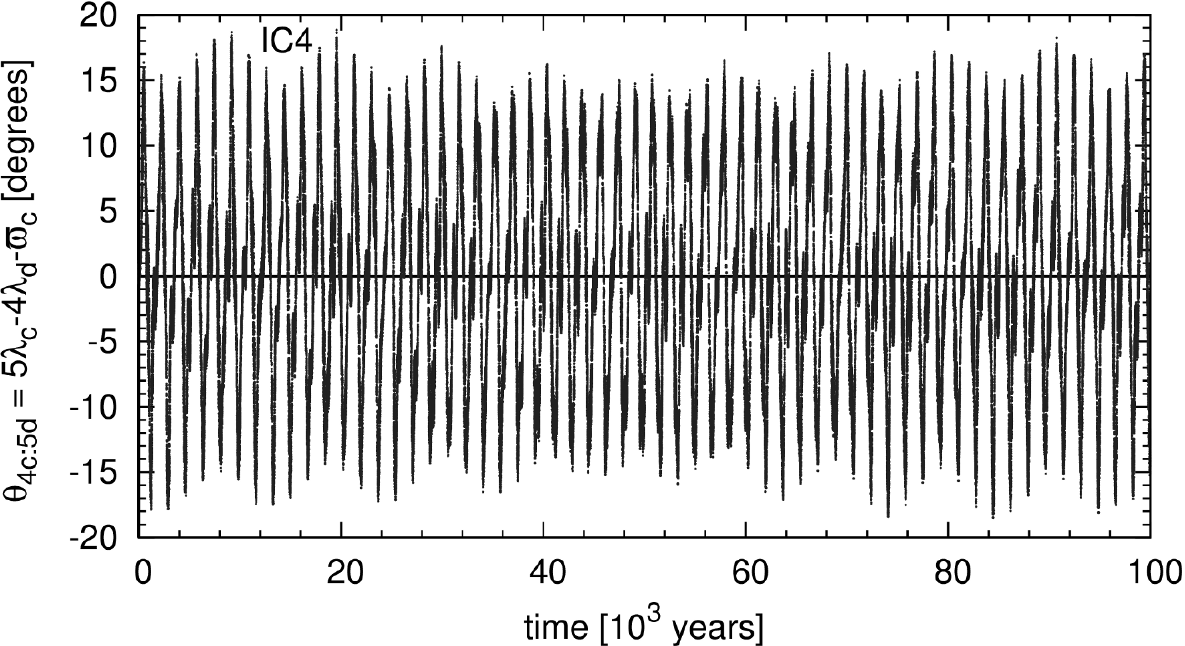}
}
}
\vskip 15pt
\vbox{
\centerline{
\includegraphics[width=0.5\textwidth]{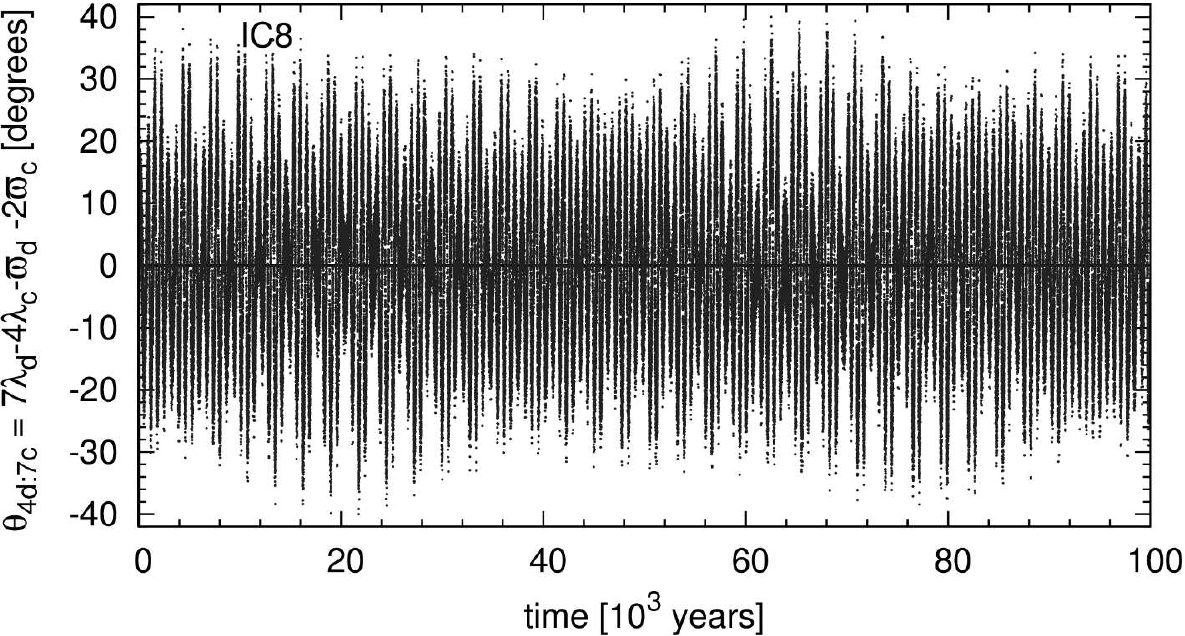}
}
}
\caption{Time evolution of the resonant angle $(\theta)$ for a selection of initial conditions (IC1, IC3, IC4 and IC8). In all cases we find the resonant 
argument to librate around zero. This finding supports the results obtained from MEGNO confirming the correct identification of quasi-periodic MMRs in the semi-major axis -- eccentricity space of the third planet.}
\label{MMRPlots}
\end{figure*}

{\bf In figure 2, we have labelled the islands of quasi-periodic orbits with their associated mean-motion resonances. 
To illustrate the true resonant character of these orbits, we have calculated their critical arguments (resonant angle) 
using

\begin{equation}
\theta_{md:lc} = l\lambda_{\textnormal{c}}-m\lambda_{\textnormal{d}} - n\varpi_{\textnormal{c}} - p\varpi_{\textnormal{d}}\,.
\end{equation}
\noindent
In this equation, $\lambda$ and $\varpi$ denote the mean longitude and longitude of pericenter, respectively and coefficients 
$(l,m,n,p)$ are integer numbers satisfying d'Alambert rule. We found that the resonant angle corresponding to each quasi-periodic 
island exhibits a clear librational 
behavior around zero degrees. We show the time evolution of the resonant angle for a few selected initial conditions (IC1, IC3, 
IC4 in the region between Kepler-47b and c, and IC8 beyond the orbit of Kepler-47c) in figure~\ref{MMRPlots}. As shown in this
figure, the variation of the resonance angle for IC8 has larger amplitudes. The MMR-locking is deep in all cases and supports the 
interpretation that the islands of quasi-periodic orbits in the MEGNO dynamical maps are associated with mean-motion resonances. 
This results also has encouraging implications for the possible detection of a third planet since MMRs can lead to an amplified 
TTV signals compared to those of the orbits in non-resonant configurations \citep{Agol05,Agol07,Haghighipour11}.}

\subsection{Predicting the Mass of the third planet}

Results shown in the previous section enabled us to identify regions of the parameter space where a test particle could 
maintain a stable orbit alongside the two massive planets, Kepler-47b and c. In reality, however, the third planet is 
also massive and its interaction with other planets can alter the orbital architecture of the system. In the following, 
we will explore the dynamical characteristics of the system by considering the third planet to be massive as well. 

As stated by \cite{Orosz2012b}, the mass of the putative third planet is practically unconstrained. However, 
the depth of its single transit points towards a planet with a radius of $\simeq 5$ Earth-radii. 
Because of the lack of a precise model for the density and interior of such planets, we consider a range of 
values for the mass of the third planet from 0 to 150 Earth-masses. 

\begin{figure*}
\begin{center}
\vbox{
\centerline{
\includegraphics[scale=0.7]{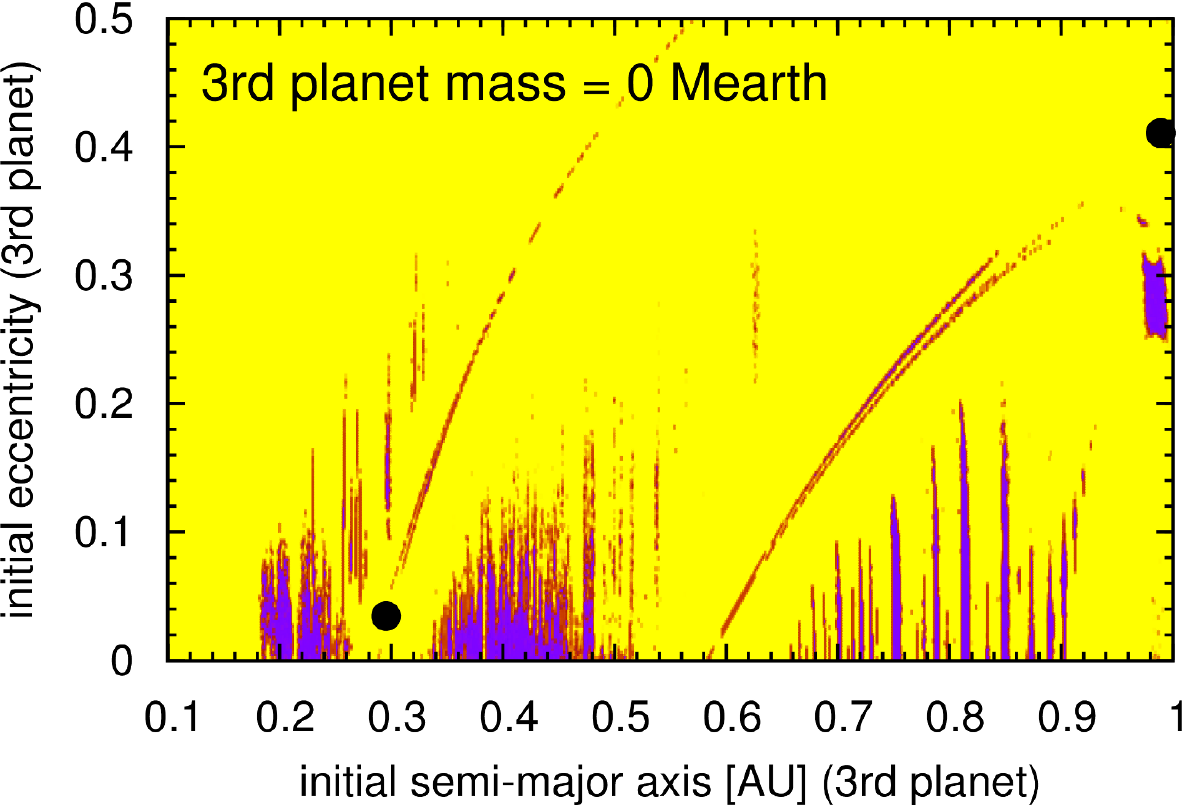} 
\includegraphics[scale=0.7]{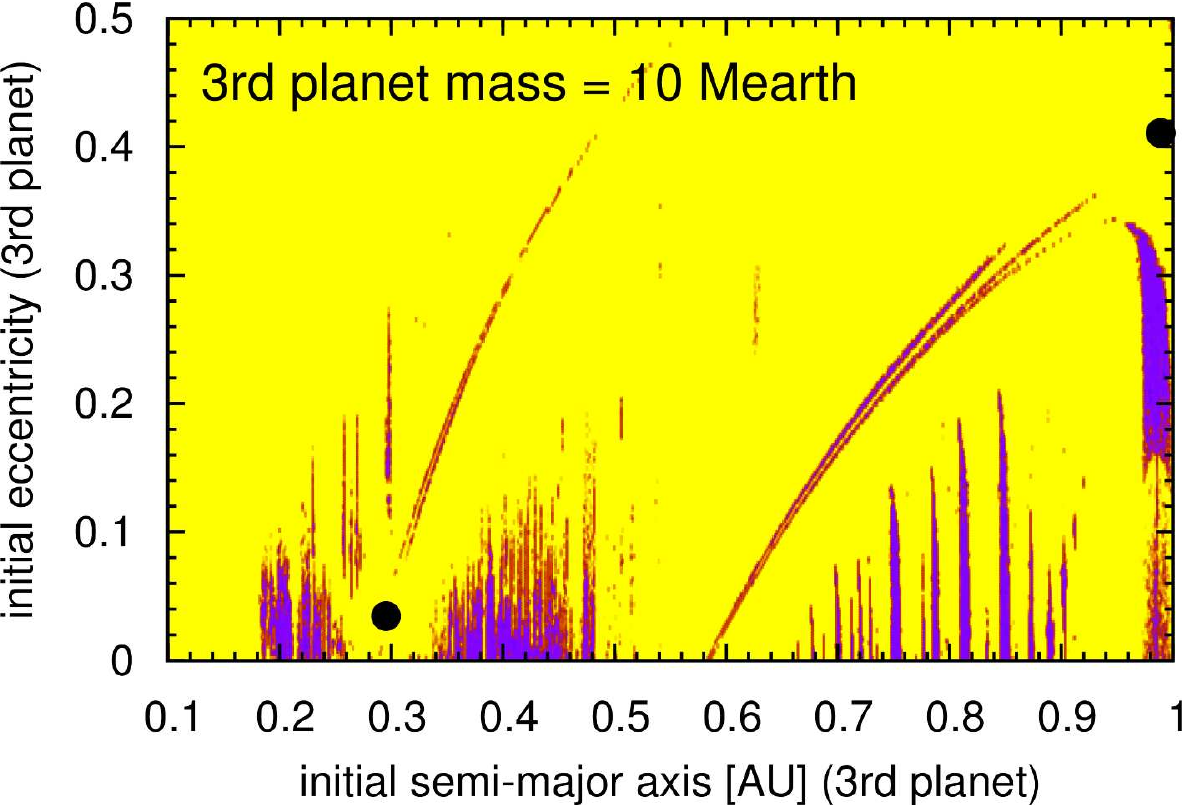} 
}
}
\vbox{
\centerline{
\includegraphics[scale=0.7]{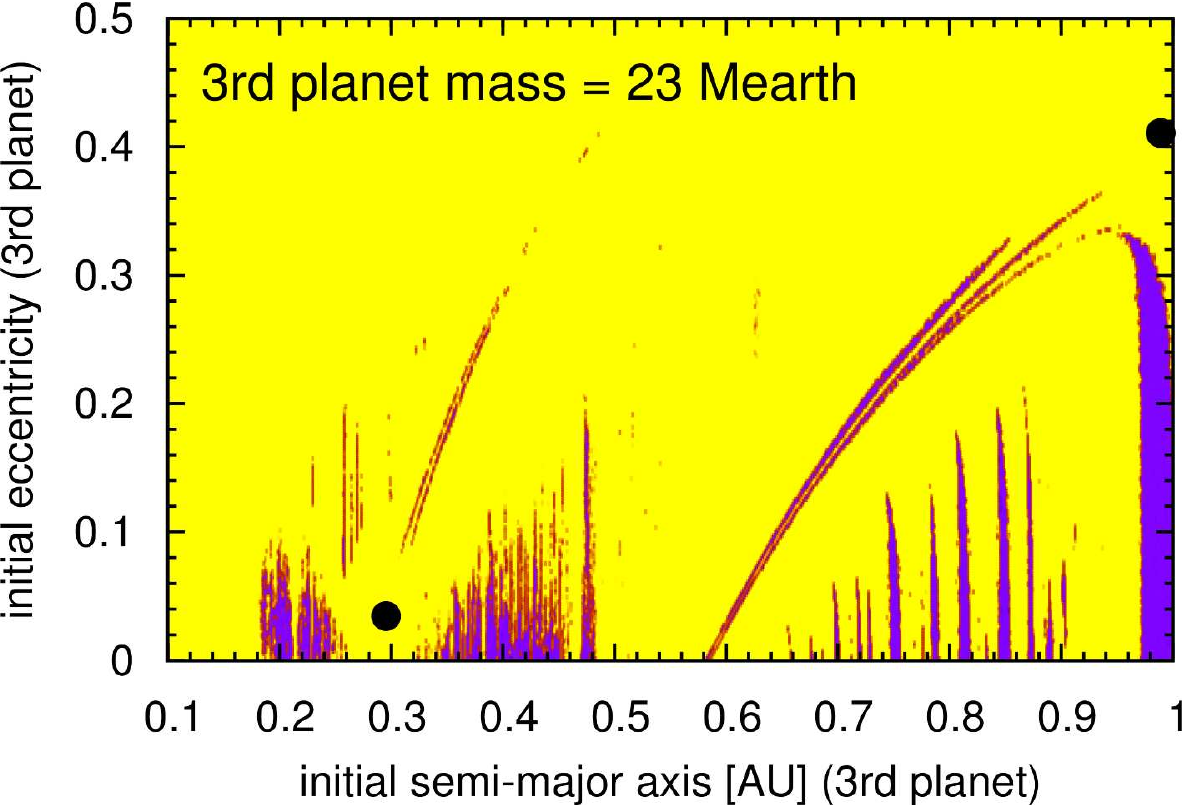} 
\includegraphics[scale=0.7]{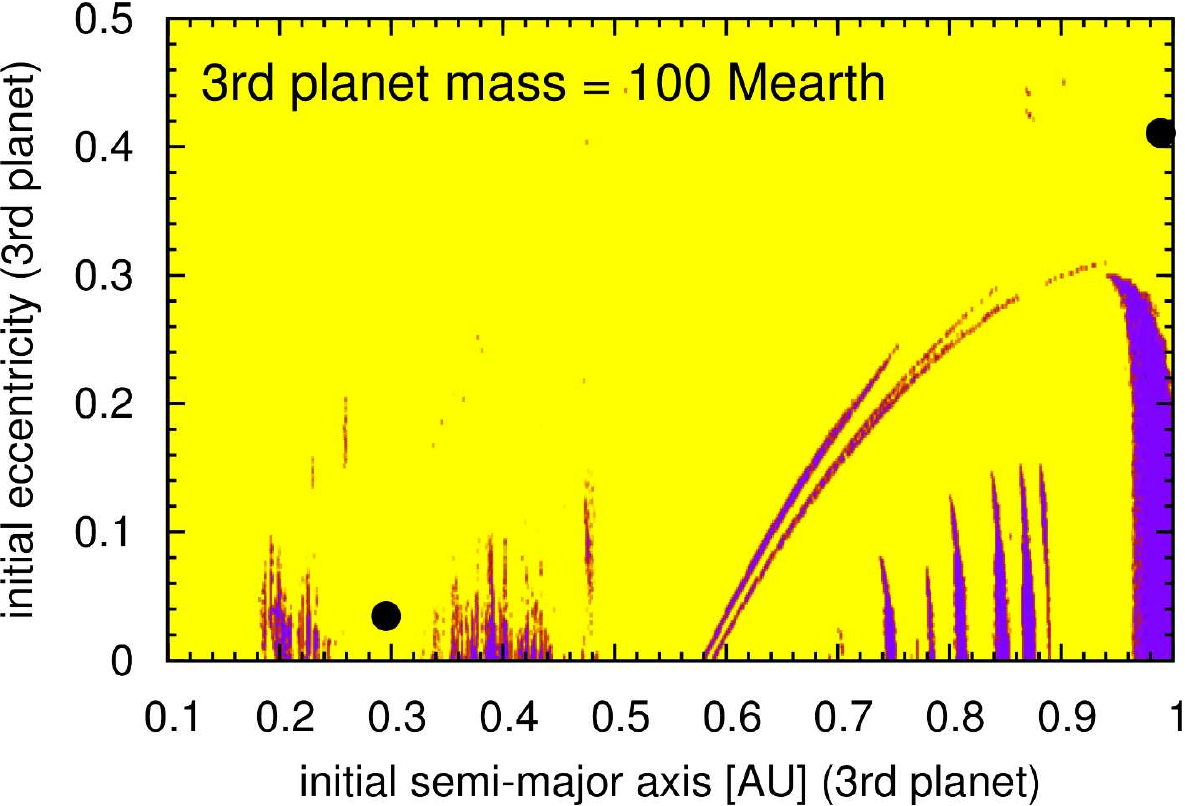} 
}
}
\caption{Dynamical MEGNO maps of the third planet for different values of its mass. Integrations were carried 
out considering the full five-body system. The two planets Kepler-47b and c are shown by black circles. Initial 
orbital elements of these two planets were taken from Table \ref{planetparams}. Color coding is the same as in 
figure ~\ref{K47_Map017}.}
\label{masssurvey1}
\end{center}
\end{figure*}

\begin{figure*}
\begin{center}
\vbox{
\centerline{
\includegraphics[scale=0.7]{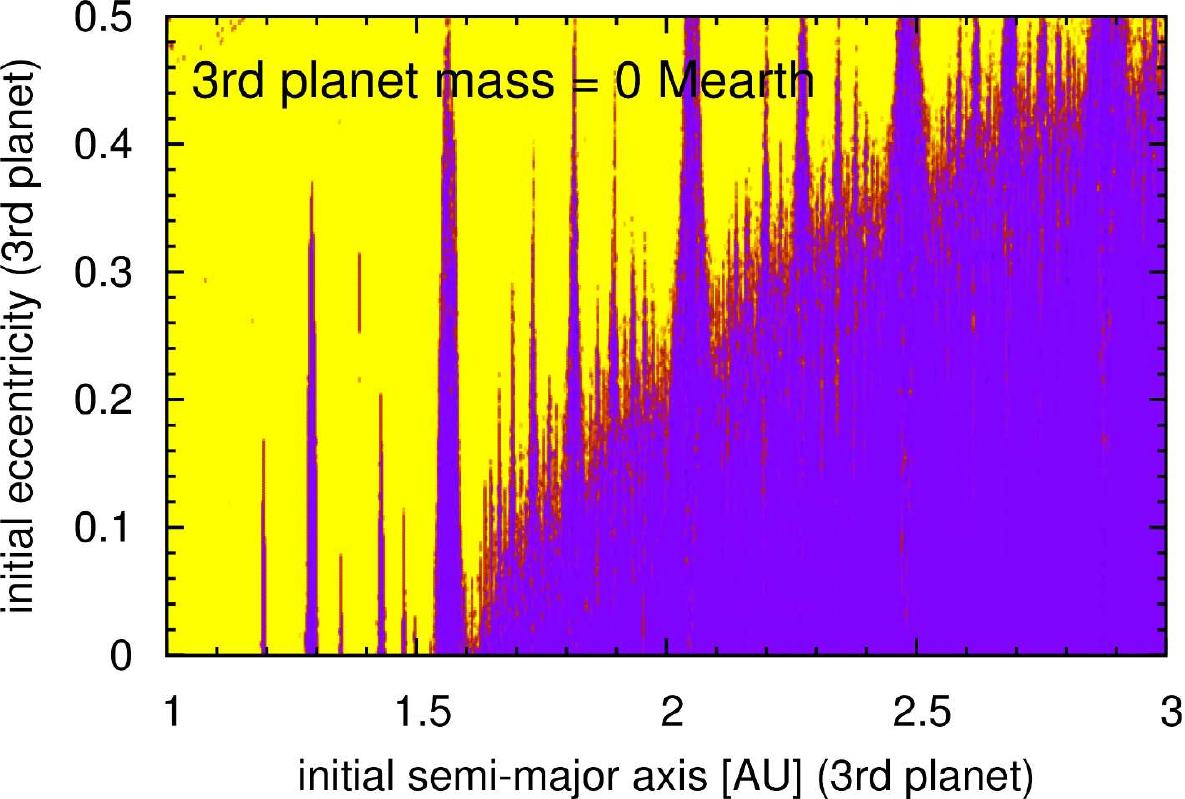} 
\includegraphics[scale=0.7]{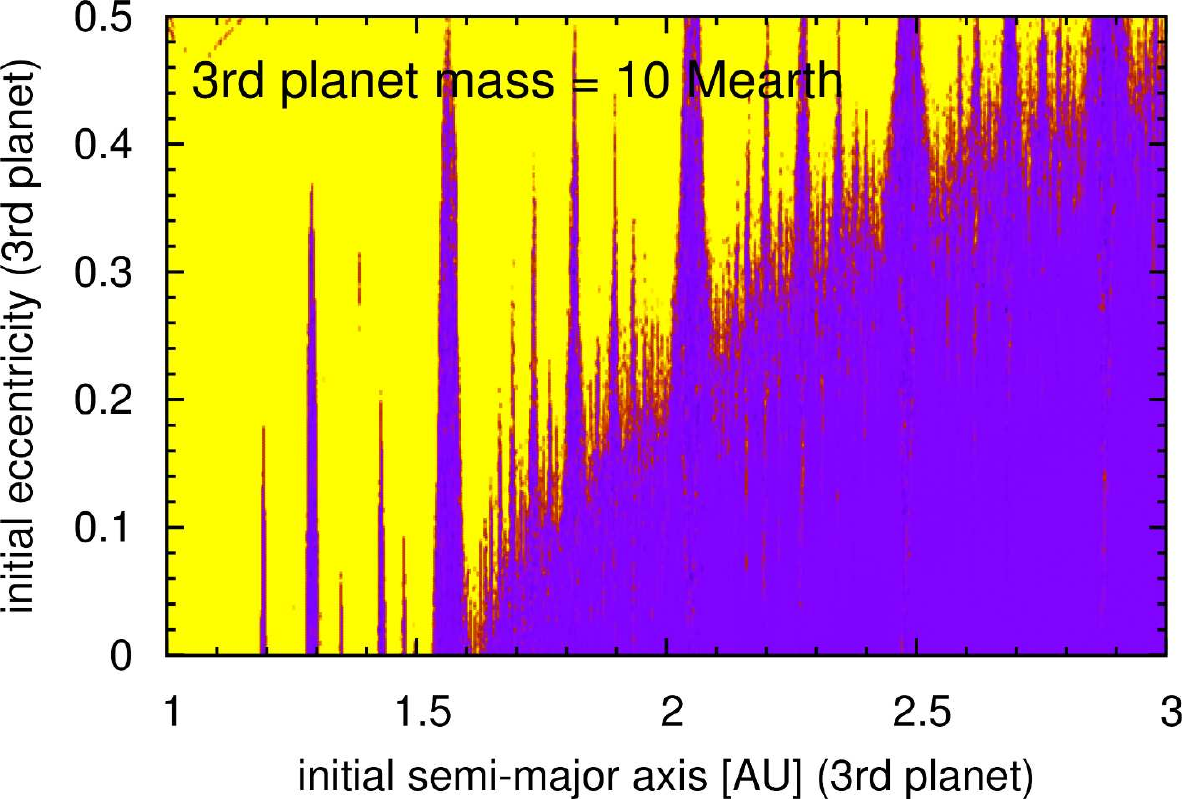} 
}
}
\vbox{
\centerline{
\includegraphics[scale=0.7]{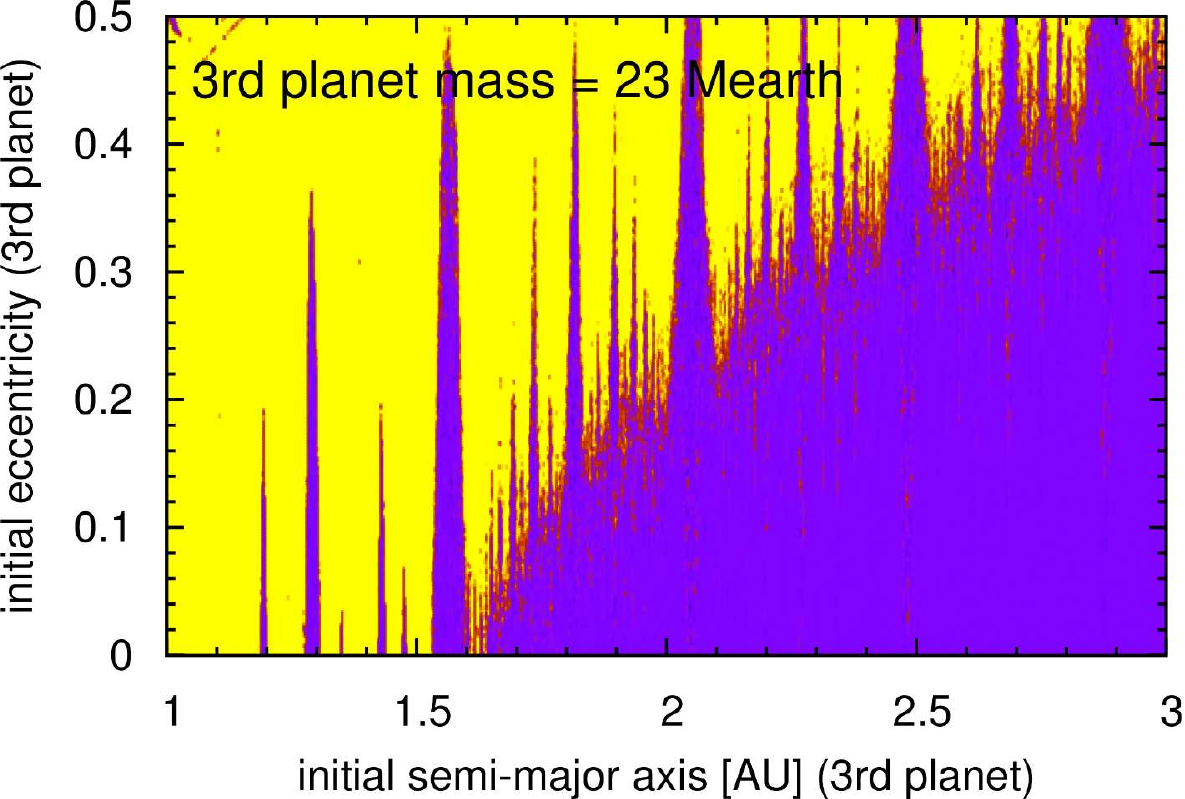} 
\includegraphics[scale=0.7]{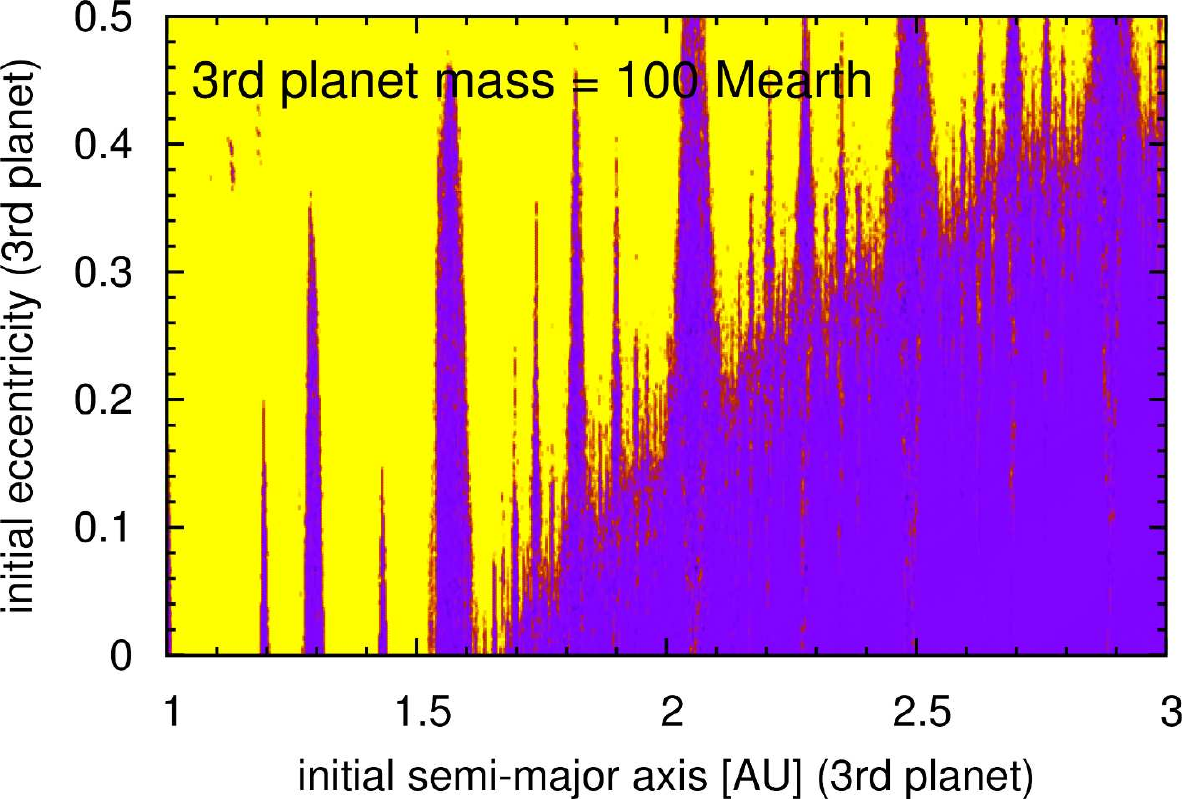} 
}
}
\caption{Same as  figure ~\ref{masssurvey1} but probing the region exterior to the orbit of Kepler-47c. 
Color coding is the same as in figure ~\ref{K47_Map017}.}
\label{masssurvey2}
\end{center}
\end{figure*}

Figures ~\ref{masssurvey1} and \ref{masssurvey2} show the results of the calculation of MEGNO  
for the entire 5-body system considering the third planet to have a mass of 0, 10, 23 and 100 Earth-masses. 
We probed the semimajor axis of the third planet in the interval of 0.1 AU to 1 AU in figure ~\ref{masssurvey1} 
and between 1 AU and 3 AU in figure \ref{masssurvey2}. In all maps, we considered the orbital eccentricity 
of the third planet to vary in the range of 0 to 0.5. As shown in these figures, there are regions between
planets Kepler-47b and c where the third planet can have stable orbits. These figures also show the change in the mass of 
the third planet does not seem to have a significant impact on the overall dynamical structure of the system. 
However, subtle differences do appear. For instance, a close inspection of the
results indicates that the quasi-periodic regions around Kepler-47b tend to diminish with increasing the mass of the 
third planet. Also, considering a massless third planet at a semimajor axis of 0.989 AU with an eccentricity of 0.15 seems to  
result in a chaotic orbit. However, increasing the mass of the planet to 100 Earth-masses renders the orbit with the same 
initial condition, quasi-periodic.

We also noticed that some of the mean-motion resonances around 0.8 AU had vanished when considering the third 
planet to be of 100 Earth-masses. For such a high value of the mass, the locations of quasi-periodic mean-motion 
resonances between the third planet and Kepler-47c appear to have shifted to smaller semimajor axes. The effect of 
changing the values of the mass of a planet on the location of mean-motion resonances has 
been illustrated in detail by \cite{Gozdziewski2013} in their study of the system of $\nu$ Octantis. In general, resonances 
may become wider/narrower or even split if the masses of planets change. Diminished quasi-periodic regions exterior to the 
orbit of Kepler-47c were also observed when the third planet was taken to be 100 Earth-masses. It seems that 
low-eccentricity, co-orbital, quasi-periodic orbits are more likely when the third planet has a higher mass.

Results shown in figures ~\ref{masssurvey1} and \ref{masssurvey2} suggest that a third planet in a 
low-eccentricity orbit can be stable in the region between Kepler-47b and c, and also for semimajor axes larger than 1.75 AU. 
The main requirement for long-term orbital stability is set by the third planet's pericenter distance. In order for the 
planet to be stable, this distance has to be well-separated from Kepler-47b and c (hence its low-eccentricity orbit) 
so that the gravitational perturbation of these bodies cannot alter the dynamics of the planet.

\begin{figure*}
\vbox{
\centerline{
\includegraphics[width=0.5\textwidth]{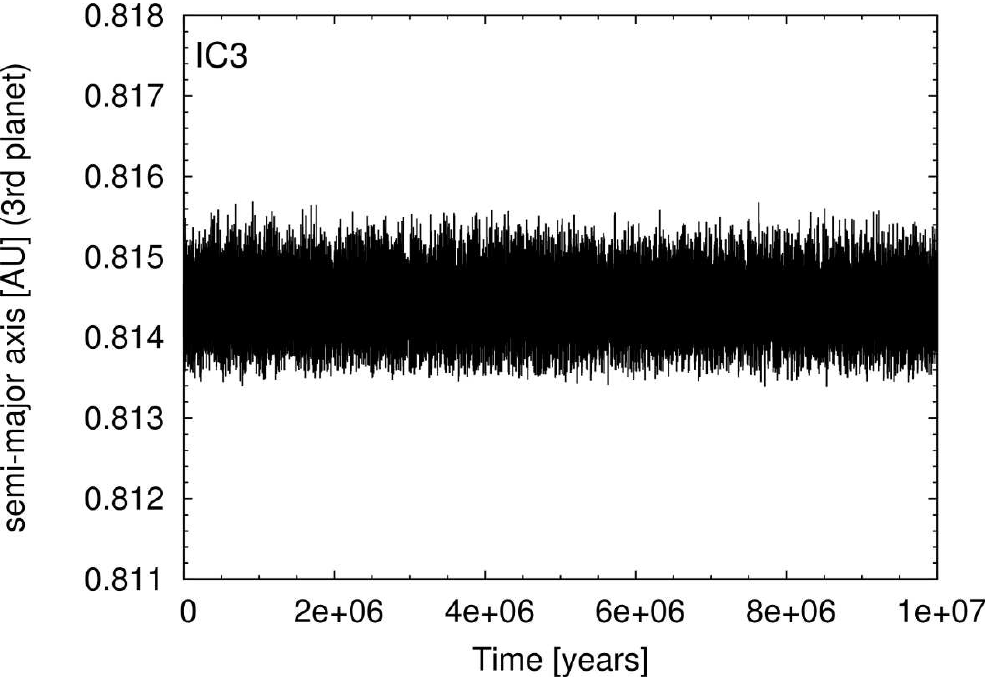}
\includegraphics[width=0.5\textwidth]{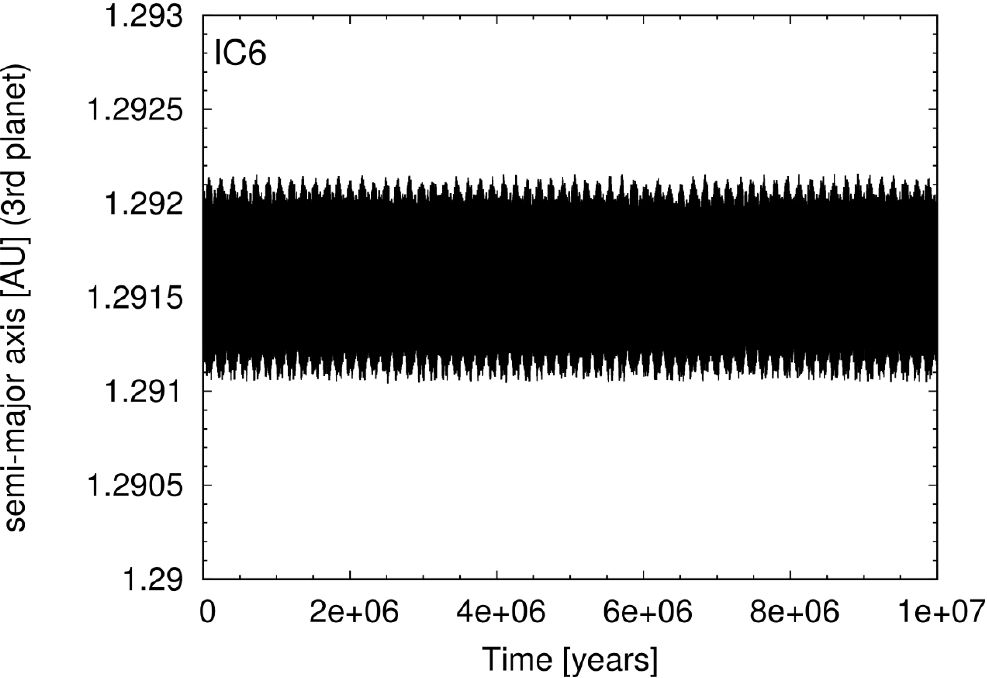}
}
}
\vskip 15pt
\vbox{
\centerline{
\includegraphics[width=0.5\textwidth]{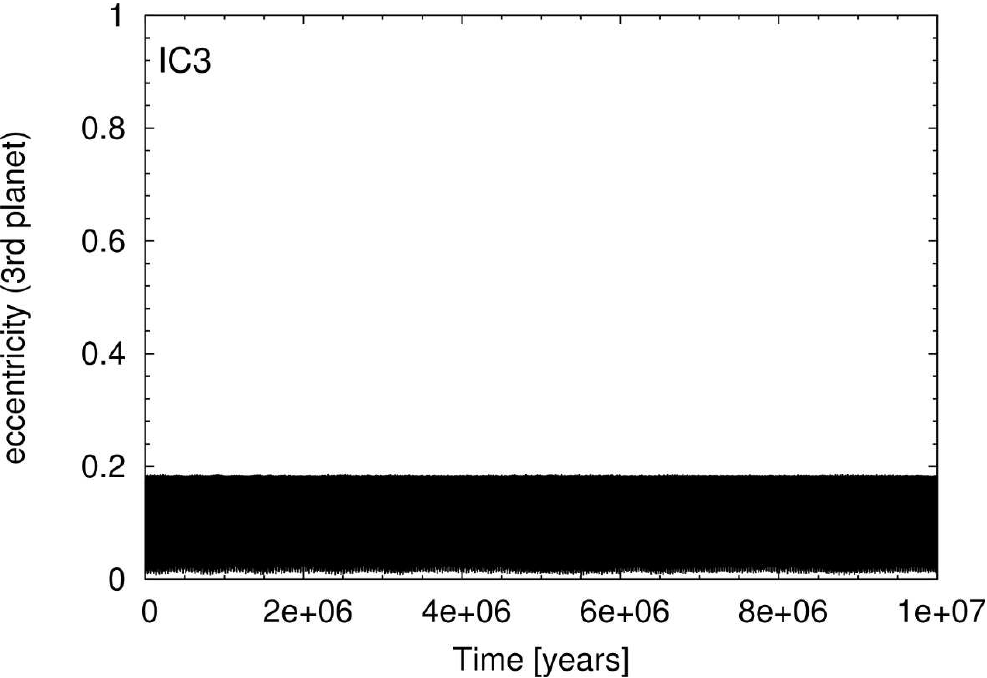}
\includegraphics[width=0.5\textwidth]{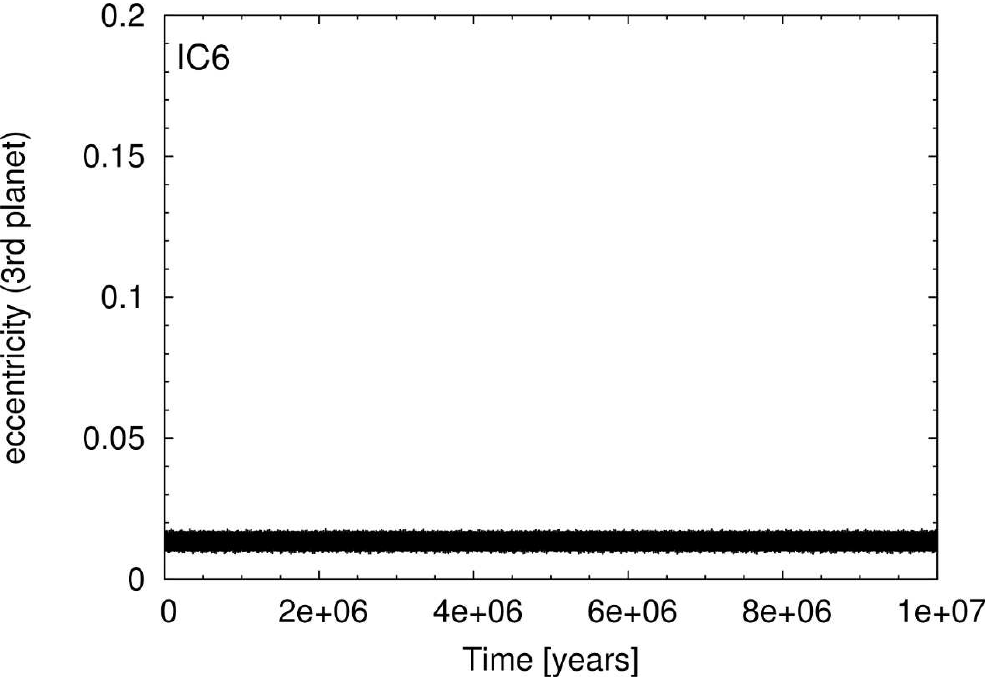}
}
}
\vskip 15pt
\vbox{
\centerline{
\includegraphics[width=0.5\textwidth]{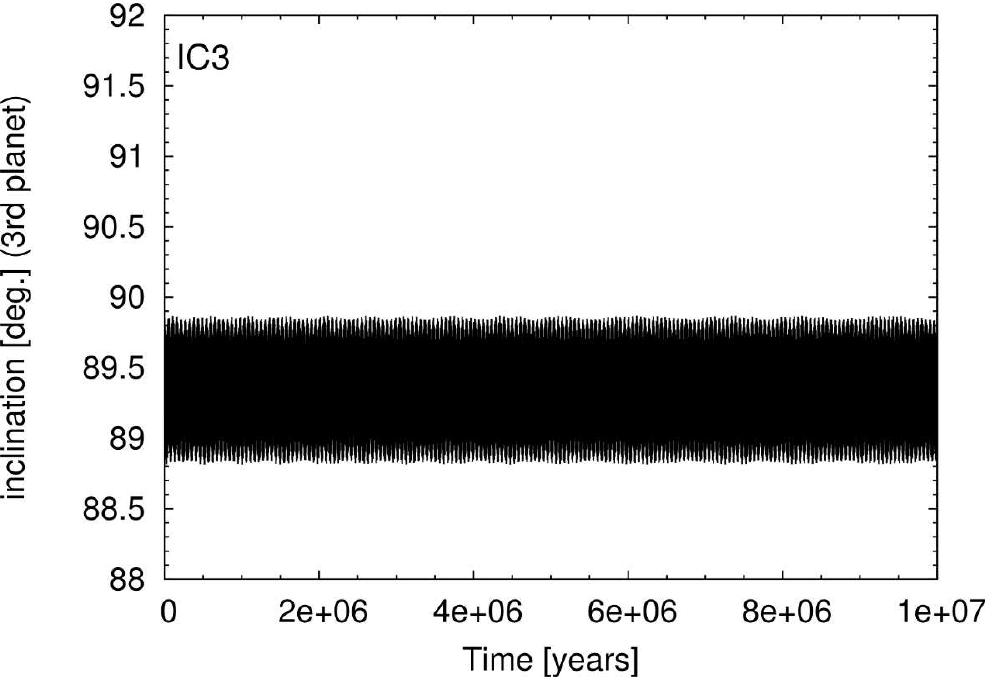}
\includegraphics[width=0.5\textwidth]{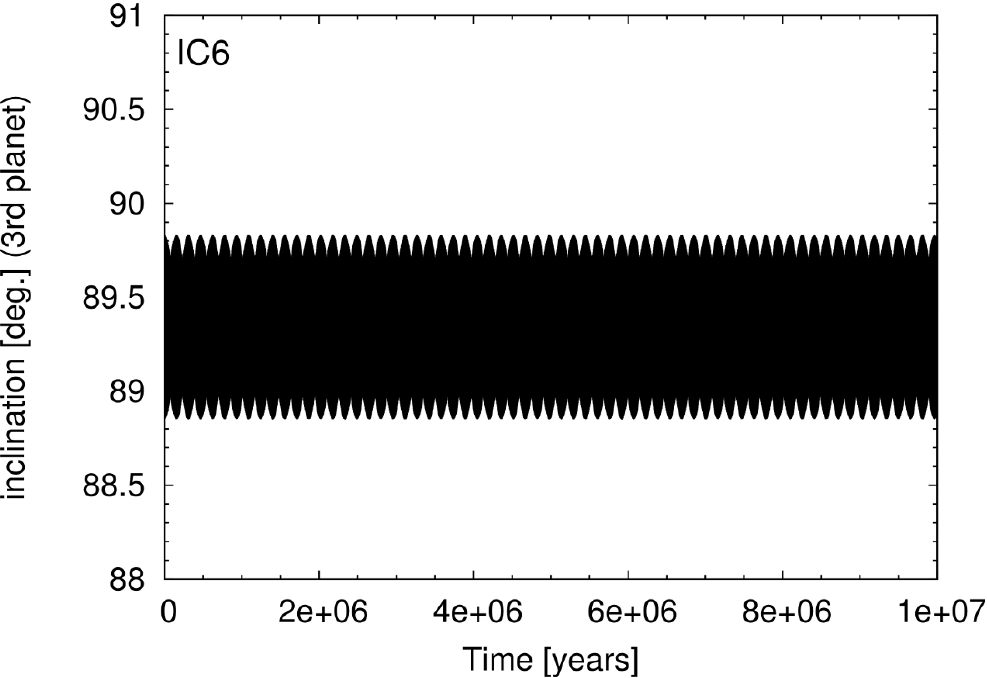}
}
}
\caption{Time evolution of (Jacobian) orbital element of the third planet considering initial conditions IC3 (left; 
third planet between Kepler-47b and c) and IC6 (right; third planet exterior to Kepler-47c). In both cases, the mass
of the third planet is 50 Earth-masses.}
\label{K47_Orbit22E_Orbit25Orphan50Mearth}
\end{figure*}

To examine the long-term stability of the third planet, we carried out two single-orbit integrations for $10^7$ years. 
We considered the third planet to be 50 Earth-masses and chose its initial semimajor axis and eccentricity the same as in
IC3 in figure ~\ref{K47_Map055_Map056} (the third planet being interior to Kepler-47c) and IC6 in 
figure ~\ref{K47_Map055_Map056} (the third planet being exterior to Kepler-47c). Figure ~\ref{K47_Orbit22E_Orbit25Orphan50Mearth} 
shows the result. As shown here, no sign of chaotic diffusion is observed for any values of the orbital elements 
of the third planet. This suggests that a third planet with a mass of 30 Earth-masses or smaller can have a stable 
orbit either between the two known planets (IC3) or in an orbit exterior to Kepler-47c (IC6). The orbits of planets 
b and c exhibit a very weak signature of chaotic dynamics. In both simulations, the orbits of these planets remained
bound and did not show any sign of a random walk.

We carried out similar simulations for higher values of the mass of the third planet. We found that the simulation 
with initial conditions IC3 results in chaotic and unstable orbits when the mass of the third planet is as high as 
150 Earth-masses. However, when starting the third planet 
with initial conditions IC6, we found that larger masses are allowed rendering an overall stable configuration. 
 
\section{Analysis of transit timing and transit duration variations}

As shown in the previous section, dynamical considerations point to regions in the Kepler-47 two-planet system
where a third body can have a stable orbit. However, this orbit is unconstrained. In this section, we use the measured
time and duration of transits in the Kepler-47 system to constrain the orbit of the third planet. 

We start by calculating transit duration (TDV) and transit timing variations (TTV) induced by the third planet on Kepler-47b. 
Due to its shorter orbital period, this planet has higher number of transits providing a large set of 
measurements for comparison and verification purposes. 

To calculate the values of TTV and TDV of a planet, we use the same numerical integration algorithm that was used in generating
MEGNO maps. We integrate the full five-body system using the values of the masses and radii of the stars and planets as well as 
the orbital elements of the two known planets as given in Table 1.
During the integration, we monitor the on-sky projected position of the planet ($r_{\rm sky}$) 
relative to the primary star. Once the planet has crosses the north-south axis passing through the center of the star, we determine the 
time of ingress ($t_{1}$) from a series of back- and forth-integrations. At each integration step, the quantity $r_{\rm sky}$ is compared 
to the sum of the star and planet radii $R_{\rm A}+R_{\rm pl}$. In each iteration, our algorithm decreases the time step by a third in 
order to ensure convergence. The time of ingress is defined when $r_{\rm sky} - (R_{\rm A}+R_{\rm pl}) < 10^{-15}$. From reversing the 
velocities of all bodies, the egress time $t_{2}$ is determined. The mid-transit time is then calculated from $t_1 + (t_2 - t_1)/2$. 
We use linear regression to calculate the variations in the transit times (TTV). The transit duration (TDV) is calculated using 
$(t_2 - t_1)$. We have tested the calculation of TTV by reproducing the results given in \cite{Nesvorny2008}.

\begin{figure}
\center
\includegraphics[scale=0.60]{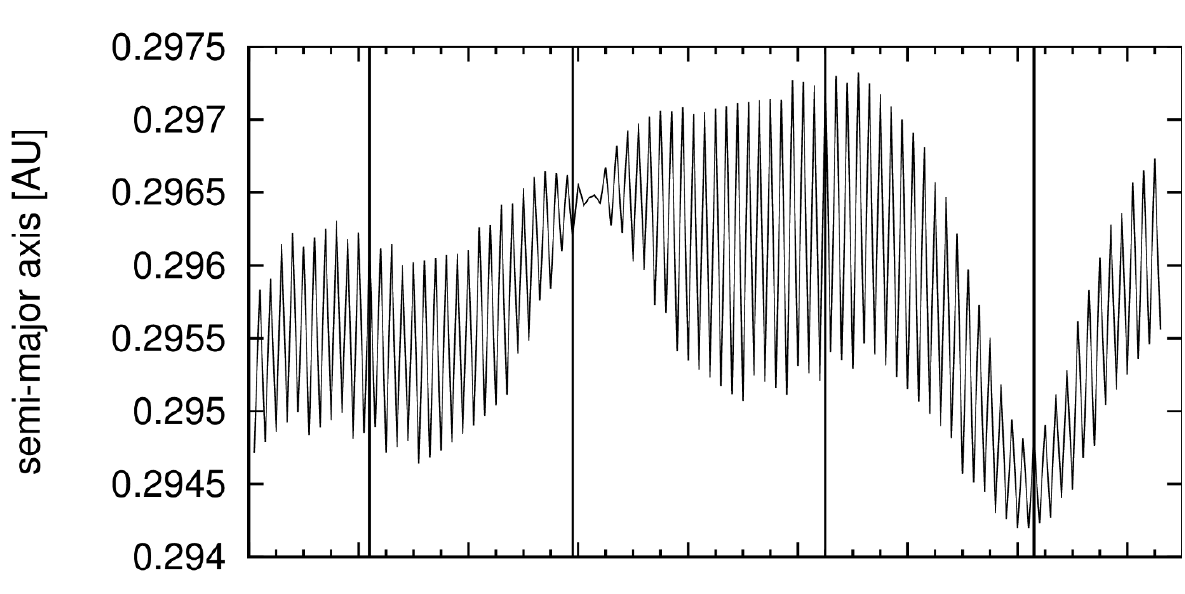}
\vskip -0.5pt
\includegraphics[scale=0.60]{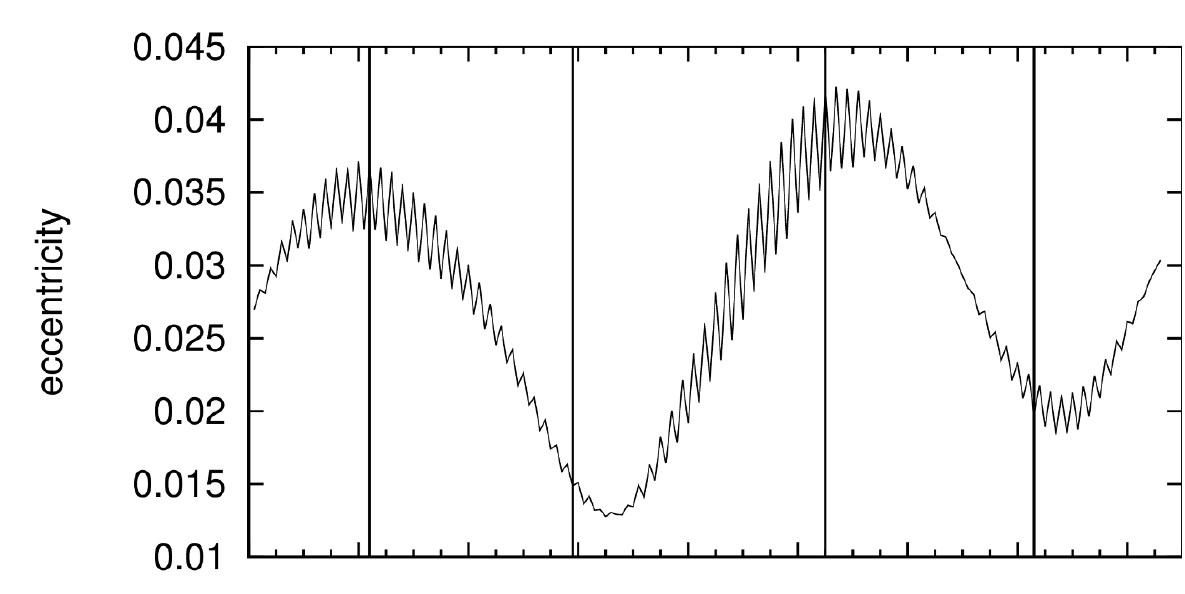}
\vskip -0.5pt
\includegraphics[scale=0.60]{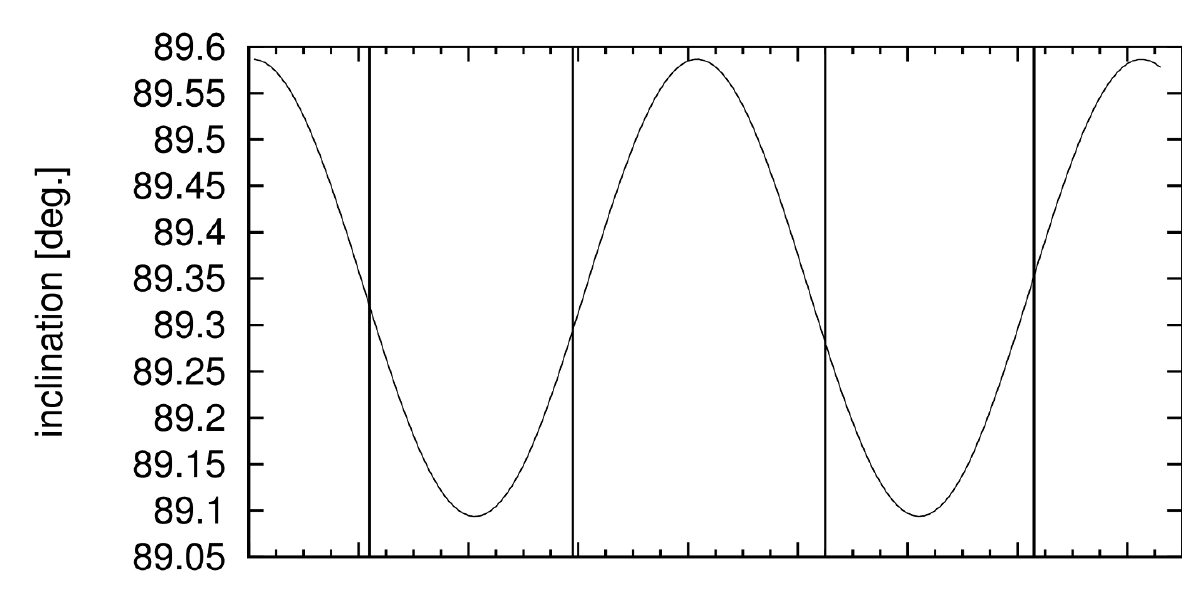}
\vskip -0.5pt
\includegraphics[scale=0.60]{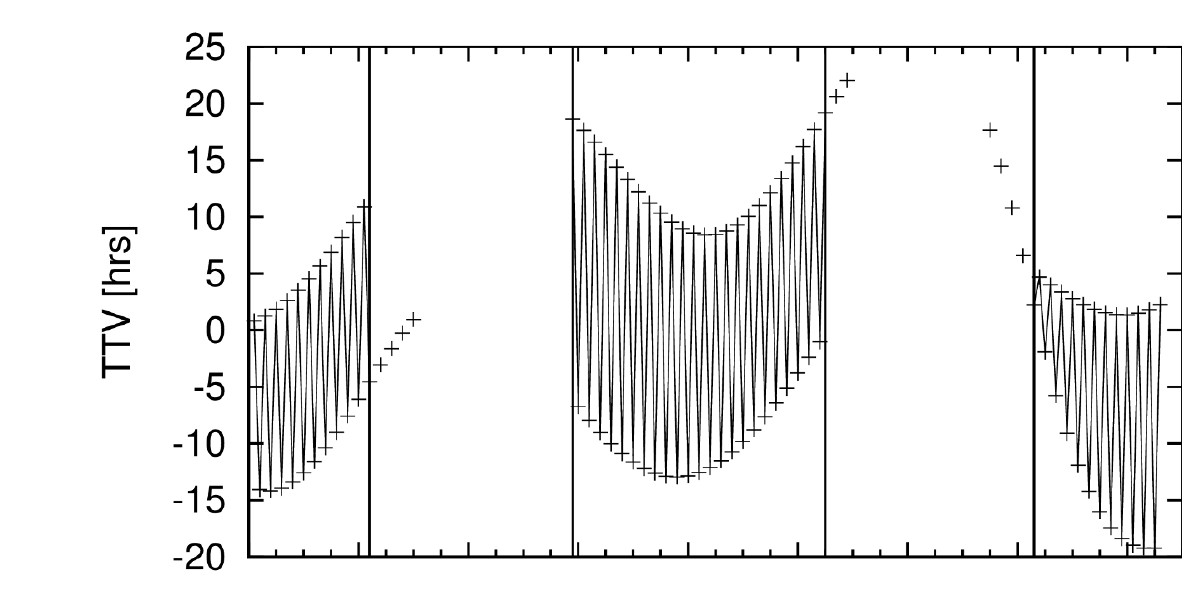}
\vskip -0.5pt
\includegraphics[scale=0.60]{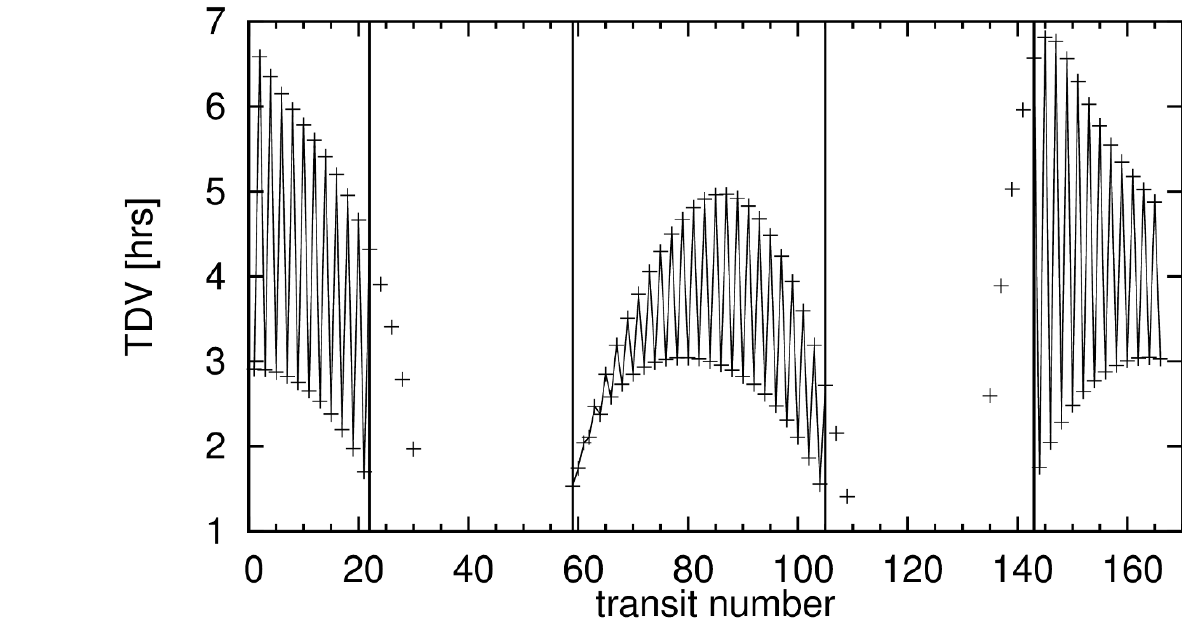}
\caption{Time evolution of the orbital elements of Kepler-47b (top three panels) and graphs of its TTV and TDV 
(bottom two panels). We considered the third planet to be Earth-massed and started it at initial condition IC6.
Vertical lines correspond to transit cycles 22, 59, 105 and 143. We note that the intersection of vertical lines 
with the inclination curve occur at different orbital inclinations as a consequence of the short-term orbital 
variations of Kepler-47b. Usually a single critical inclination is determined which corresponds to an impact 
parameter of one solar radius. The total duration corresponds to $\simeq 165\times 49.5$ days = 22.4 years.}
\label{3Planets_IC25_Orphan1Mearth}
\end{figure}

Figure ~\ref{3Planets_IC25_Orphan1Mearth} shows the results for Kepler-47b.
We considered the third planet to have 1 Earth-mass and started it at the initial condition IC6. The top three panels 
in figure ~\ref{3Planets_IC25_Orphan1Mearth} show the resulted variations in the semimajor axis, eccentricity, 
and inclination. The bottom two panels show the values of the TTV and TDV. Here we have connected 
consecutive transit events with a solid line. The vertical lines show the cycle numbers for which the first or 
last transit is detected while being part of a consecutive transit series. An example of a consecutive transit 
series is the transits 59 to 105. As shown in the bottom two panels, there are two gaps between transits 22 and 59, and 
between transits 105 and 143 (where there is missing data). In these gaps, isolated timing measurements (shown by plus signs) 
represent non-consecutive transit events. In other words, for those TTV/TDV data points that are not connected, the 
planet did not transit the star before and after the given data point.
For instance, we see from figure ~\ref{3Planets_IC25_Orphan1Mearth} that there are four isolated near-miss 
transits after the transit cycle 22. There are missing transits in all gaps which have not been plotted for 
the obvious reason.
Such missed transit events were also observed in the light curve of Kepler-413b \citep{Kostov2014}. 
The only explanation for such single-transit or near-miss events is the short-term variations in the planets orbital 
elements. As shown in figure ~\ref{3Planets_IC25_Orphan1Mearth}, the orbital elements 
of Kepler-47b do indeed undergo significant changes from one transit to another suggesting that
the gaps are due to the low orbital inclination of Kepler-47b relative to the plane of the sky. We  
recall that the closer the inclination is to $90^{\circ}$, the closer the planet will be transiting along the star's 
equator. Therefore, for the specific geometry of the Kepler-47 system, 
the projected orbital plane of Kepler-47b on the plane of the sky will be outside 
the star's disk $(r_{\rm sky}-R_{\rm pl} > R_{A})$ and therefore, no timing measurements can be computed. 
As a result, such events do not appear in figure ~\ref{3Planets_IC25_Orphan1Mearth}. 
However, for a given value of the orbital inclination, we do obtain single transit events and alternating 
missing transits when Kepler-47b is close to the edge of the star. 

Interestingly, we see no near-miss or isolated transit prior to transit 59. A detailed examination of the light curve at around transit 59 shows that the transit-to-transit variations in semimajor axis and eccentricity of Kepler-47b are minimal. We conclude that the short-term orbital variations are significant 
and capable of shifting the planet's projected disk in and out of the stellar disk. This is an interesting result as missing transit events can be used to further constrain the results of photodynamical models in future discoveries of transiting circumbinary planets. 


\subsection{Transit duration of the third planet}

In this section, we calculate the durations of the transits of the third planet and compare them with the duration
of the single, unexplained transit event of the system to constrain the orbit of this body. It is important to note that
because of the mutual interactions between planets, the duration of the transits of the third planet will vary from one
transit cycle to another. Figure ~\ref{VariousTDVs} shows this for different values of mass and initial orbital 
configuration of the third planet. In making this figure, we only considered initial conditions for which our MEGNO 
calculations indicated quasi-periodic orbits. In particular, we considered initial conditions 
IC1 to IC11 shown in figure ~\ref{K47_Map055_Map056} (except for IC2 for which the orbit of the planet was found to be 
highly chaotic). We also considered two cases for the mass of the third planet: A mass-less object shown on the left 
panels, and a 10 Earth-mass planet depicted on the right. The methodology for calculating transit duration is similar 
to the procedure described in the previous section. In each panel, the horizontal line at 4.15 hours corresponds 
to the duration of the single transit event identified by \citet{Orosz2012b}. As shown in figure ~\ref{VariousTDVs}, 
in general, the mass of 
the planet does not play a significant role in the duration of its transit. We found that in spite of its variations, 
there are several transit cycles in which the duration of the transit of the third planet is comparable to that of 
the observed single transit ($\sim 4.15$ hours). For instance, in the top-left panel of figure ~\ref{VariousTDVs} 
where the planet is mass-less and starts at the initial condition IC1, the transit number 83 appeared to have a duration 
of 4.16 hours. Transit durations of around 4 hours were also found for initial conditions IC3 and IC4. We recall 
that IC1, IC3 and IC4 are initial conditions where the third planet starts between the two planets Kepler-47b and c. 
We also found that when the planet is massive, transit durations of around 4 hours 
are possible for the initial condition IC6 where the orbit of the third planet is exterior to Kepler-47c.

The results mentioned above point to a strong degeneracy. It seems impossible to determine the correct orbit of 
the third planet using its transit duration. This degeneracy can, however, be broken assuming the third planet 
is in a circular orbit. As shown by \cite{Kostov2013}, in this case, the orbital period of the third planet 
can be determine using

\begin{equation}
P_{\rm d} = P_{\textnormal{bin}} (0.74 \sqrt{1-b^2} + 0.26)^{-3}\,,
\end{equation}

\noindent
where $b$ is the (a priori unknown) impact parameter. As $b$ varies only between 0 and 1, we can use equation (1) to 
constrain the orbital period of the third planet and break the degeneracy. Specifically, if $b=0$ (i.e. a central transit) 
then $P_{\rm d} = P_{\rm bin}$, representing a third body with the same orbit as that of the binary. This is consistent with 
the measured transit duration of 4.15 hours being comparable to the duration of the stellar eclipse. On the contrary, if 
$b \approx 1$ (i.e. a grazing transit), equation (1) gives $P_{\rm d} = 57 \times P_{\rm bin}$, suggesting an upper limit 
for the period of the third planet of $\sim 424$ days. For a binary mass of $1.4 {M_\odot}$ this corresponds to a semimajor 
axis of $\sim 1.24$ AU, effectively ruling out initial conditions IC6 through IC11, as well as orbits with progressively 
larger semimajor axis. Thus the third planet is either on a stable orbit in the vicinity of Kepler-47b or between the two known planets Kepler-47b and c, or started along IC5.

\section{Discussion \& Conclusions}

In this study, we used dynamical considerations to examine  the possibility that the single, unexplained transit 
in the Kepler-47 system as reported by \cite{Orosz2012b} would be due to a third planet. Using the MEGNO technique, 
we identified regions in the phase-space where the third planet could follow quasi-periodic orbits considering the 
five-body problem. We determined several of such quasi-periodic regions between the two known planets Kepler-47b and 
Kepler-47c, where they also include orbital mean-motion resonances with either one of the two bodies.

Using accurate single orbit integrations, we examined the long-term orbital stability of the third planet within the 
framework of a five-body problem. Results identified ranges of semimajor axis and eccentricity that would render the 
third planet stable over a time period of 10 million years. To examine the extent of the dependence of the results 
on the mass of this planet, we carried out integrations for different values of its mass, and showed that
a third planet with a mass as high as 50 Earth-masses can still maintain stable orbits either between Kepler-47b 
and Kepler-47c, or beyond the orbit of Kepler-47c. For higher masses of the third planet the quasi-periodic stable 
regions in the vicinity of Kepler-47b ceases. For a selection of initial conditions within quasi-periodic islands 
we established a clear association of these islands with two-body mean-motion resonances by demonstrating a librating (around zero) critical argument.

To constrain the orbit of the third planet, we calculated its transit durations as well as the TTV and TDV of Kepler-47b
for various initial conditions. We found that transit duration and transit timing variations are affected by short-term 
changes in the orbit that have been caused by perturbations due to other bodies in the system. This implies that it is 
imperative to consider gravitational interactions when studying multi-body systems where mutual perturbations can be significant. 

\begin{figure*}
\vbox{
\centerline{
\includegraphics[width=0.5\textwidth]{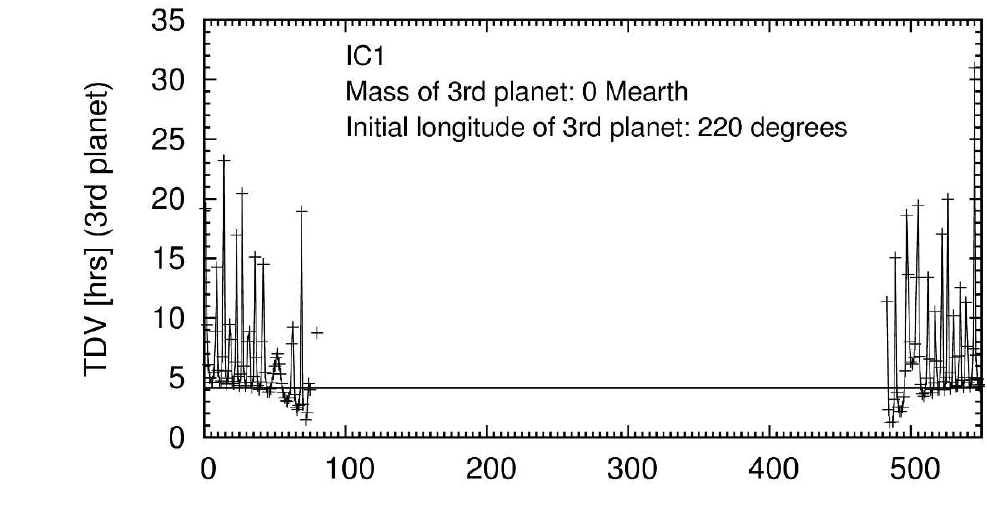}
\includegraphics[width=0.5\textwidth]{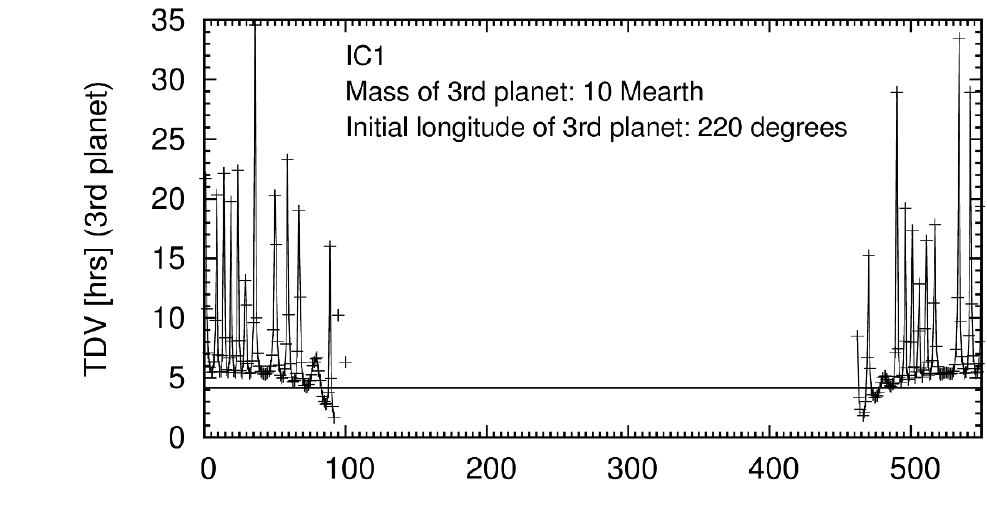}
}
}
\vbox{
\centerline{
\includegraphics[width=0.5\textwidth]{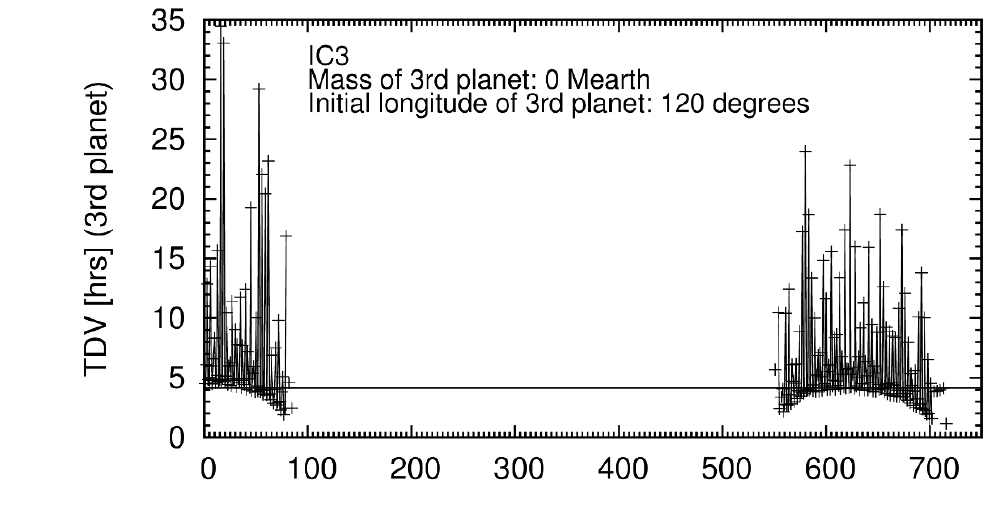}
\includegraphics[width=0.5\textwidth]{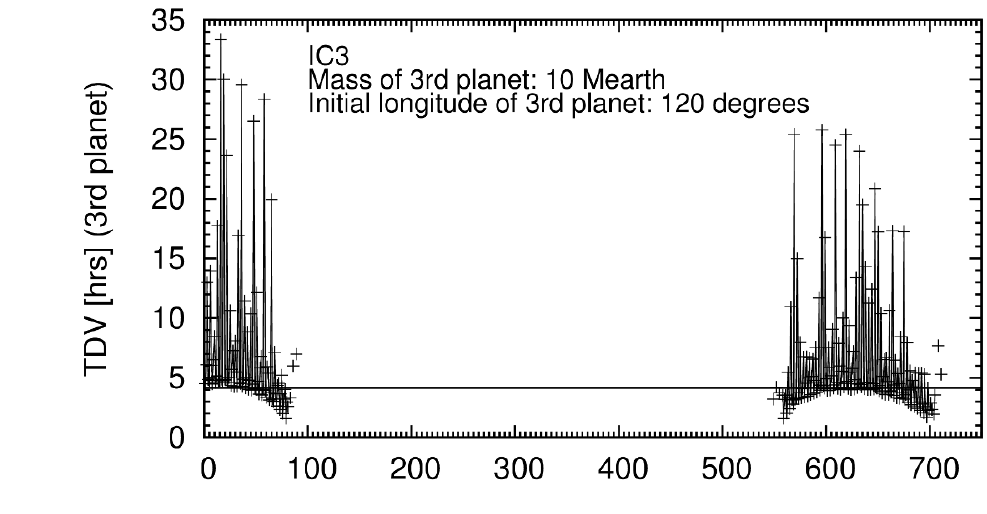}
}
}
\vbox{
\centerline{
\includegraphics[width=0.5\textwidth]{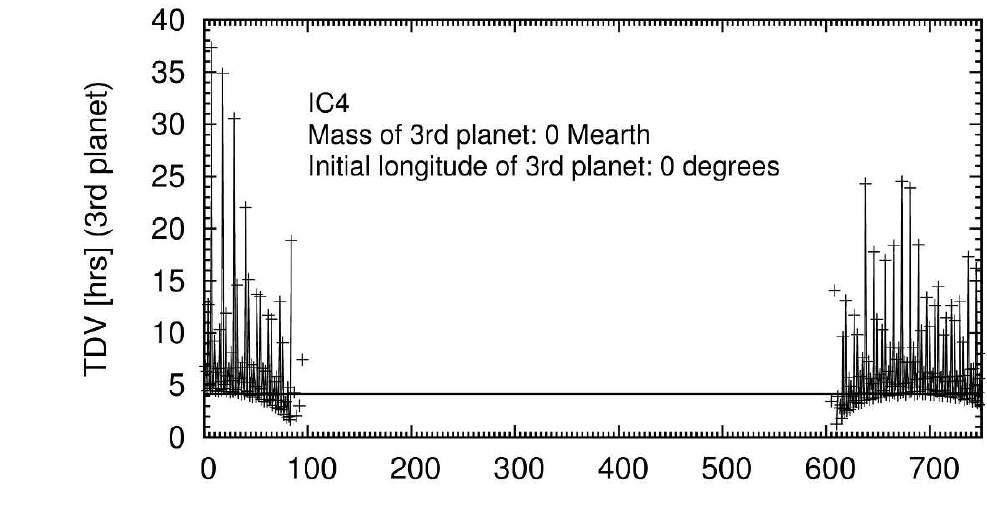}
\includegraphics[width=0.5\textwidth]{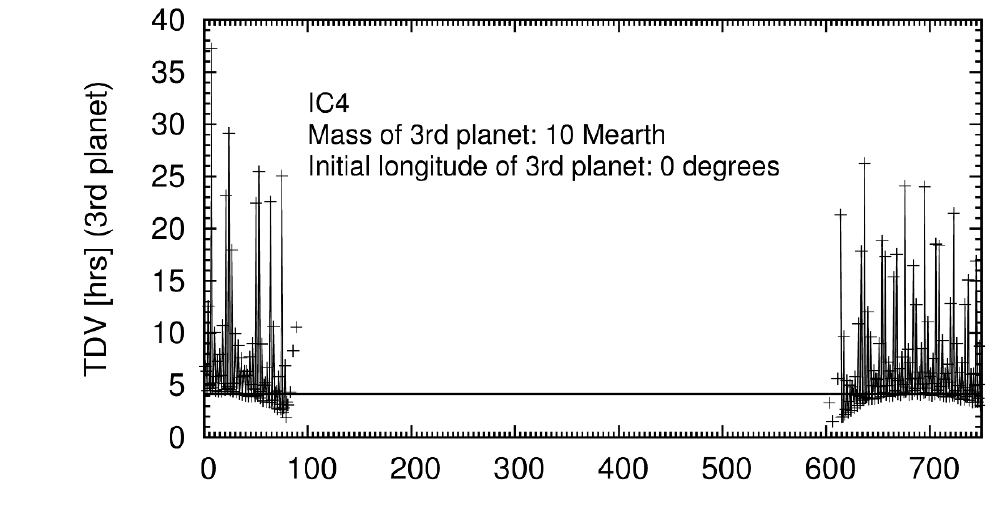}
}
}
\vbox{
\centerline{
\includegraphics[width=0.5\textwidth]{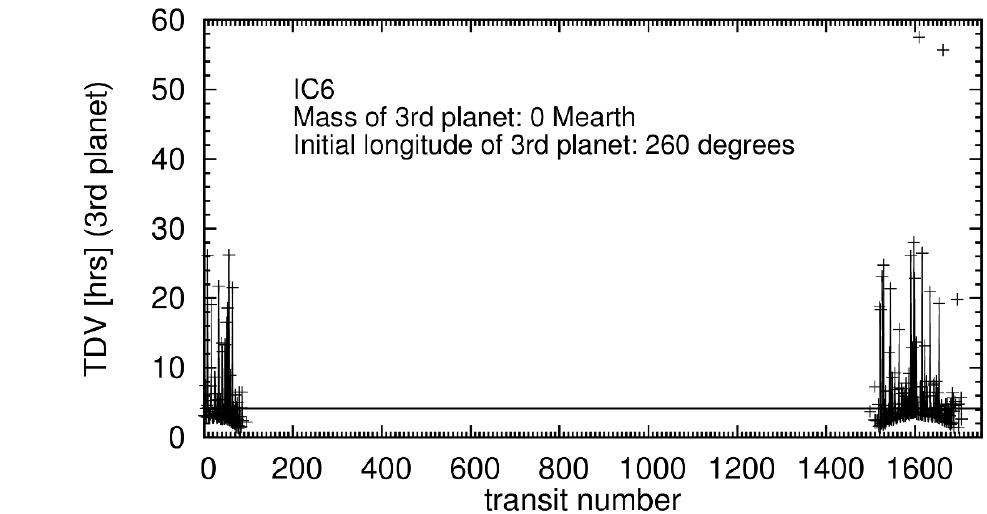}
\includegraphics[width=0.5\textwidth]{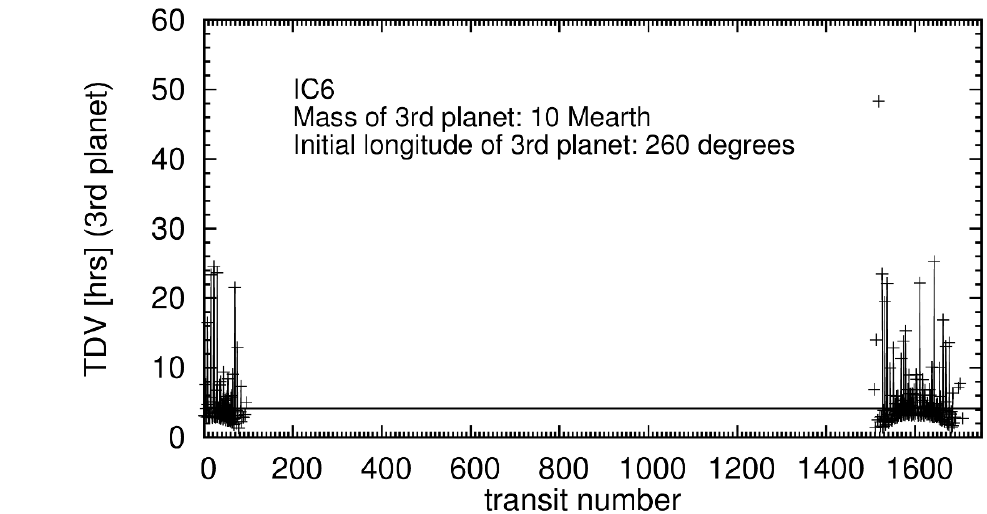}
}
}
\caption{Graphs of the transit duration of the third planet for various initial conditions. The left column corresponds to
a massless object and the right column is for a 10 Earth-mass planet. The horizontal line marks a transit duration of 4.15 hours. 
We considered various initial mean longitudes of the third planet. Other initial mean longitudes were also tried and resulted 
in similar results. As shown here, the duration of the transit does not have strong dependence on the initial phase of the 
third planet. Also, no significant differences between transit durations of a test mass and a massive planet were observed.
Symbols that are not connected with a line have a missed trailing or leading transit, and represent isolated 
transit events. Note the increase in the period of transit durations for longer periods of the transiting third planet.}
\label{VariousTDVs}
\end{figure*}

{\bf Also, when the line-of-sight inclination of the transiting planet becomes small (planet approaches the limb of the star), 
short-term perturbation on timescales on the order of the orbital period will have the effect of shifting the on-sky position 
of the planet. This can result in missing transits from one orbit to the other. These transit-missing events can be used to 
further constrain best-fit photodynamical models. In this study, for the first time, we have correlated the 
cause of missing transits with the short-term (transit-to-transit) variations of orbital elements. We also 
confirmed the possibility of long-duration transits. For instance in figure \ref{VariousTDVs}, considering the case of IC6 for a 
mass-less third planet, the duration of a single transit can last for nearly 60 hours}.

When calculating transit durations of the third planet, we found that the results suffer from a large degeneracy. 
Several orbits produced similar transit durations as that of the single transit, 4.15 hours. We were able to break 
this degeneracy for circular orbits and determined an upper limit of 424 days for the orbit of the third planet.

{\bf In a recent study by \cite{KratterShannon2014}, a period of 186 days was conjectured for the third planet in Kepler-47. 
Using Kepler's third law, we obtain a semi-major axis of 0.714 AU (relative to the binary center of mass) for the third planet, 
placing it close to the stable 5c:8d mean-motion resonance or nearby resonances (see figure 2) with Kepler-47c. 
Our work predicts several islands of quasi-periodic orbits in the neighbourhood of the 186 days orbit. 
\cite{KratterShannon2014} also carried out long-term numerical integrations of the five-body system. Their results support 
the finding that all planets maintain stable orbits over at least $2 \times 10^6$ years.}

In the present analysis we suggest that a third planet could in fact be the cause of the single, unexplained transit event reported by
\cite{Orosz2012b}. This planet will have a low eccentricity orbit either i) in the vicinity of Kepler-47b (for low masses), 
ii) between planets Kepler-47b and Kepler-47c, or iii) exterior to planet c with a semimajor axis smaller than 1.24 AU.

{\bf The detection of a third planet could follow along the route of measuring TTVs or TDVs of Kepler-47b and/or Kepler-47c as caused by this planet. However, the process of finding additional bodies using timing variations is highly degenerate and can lead to various dynamical configurations that produce the same timing signal \citep{Nesvorny2009,NesvornyBeauge2010}. We therefore suggest that the entire currently available Kepler photometric data on Kepler-47 to be analyzed and searched for additional transit signatures that can not be explained by the two known planets Kepler-47b and Kepler-47c.}

\acknowledgments

TCH acknowledges support from the Korea Astronomy and Space Science Institute (KASI) 
grants 2012-1-410-02 and 2014-1-400-06, and the Korea Research Council for Fundamental Science and Technology (KRCF).
Calculations were carried out at the SFI/HEA Irish Center for High-End Computing (ICHEC) and the KMTNet computing 
cluster in South Korea. Research at the Armagh Observatory is funded by the Department of Culture, Arts \& Leisure (DCAL). TCH would 
like to thank the Institute for Astronomy and the NASA Astrobiology Institute at the University
of Hawaii for their warm hospitality during a visit in Nov/Dec 2012 when part 
of this work was carried out. TCH would also like to thank David Nesvorn{\'y} for fruitful discussion on calculating TTV and TDV. 
NH acknowledges support from NASA ADAP grant NNX13AF20G, NASA Origins grant NNX12AQ62G, 
HST grant HST-GO-12548.06-A, and Alexander von Humboldt Foundation. Support for program HST-GO-12548.06-A was provided by NASA through a grant from the Space Telescope Science Institute, which is operated by the Association of Universities for Research in Astronomy, Incorporated, under NASA contract
NAS5-26555. VBK gratefully acknowledges support from NESSF grant NNX13AM33H. 
This work has been supported by Polish National Science Centre MAESTRO grant DEC 2012/06/A/ST9/00276 (K.G.).  
The authors would like to thank to anonymous referee for valuable suggestions.

\clearpage

\clearpage

\clearpage

\clearpage

\clearpage

\clearpage

\clearpage

\clearpage

\clearpage

\clearpage

\begin{table*}
\centering
\caption{Orbital parameters of the Kepler-47 system and their $1 \sigma$ uncertainties \citep{Orosz2012b}.}
\tablecomments{{\bf The inclination of the third planet was chosen to be equal to the inclination of Kepler-47c, and the remaining 
angles were set to zero.} Orbital elements indicated by ''($\rm fixed$)'' have been undetermined from observations. 
{\bf The mass parameter $\mu=k^2(M_{1}+M_{2}+m_{i})$ has been used for a given planet with mass $m_i$ when transforming 
elements.} Symbols $R_\oplus$ and $M_\oplus$ denote the radius and mass of Earth. {\bf The mass of the primary star (Star A) is 1.043~$M_{\odot}$.} }

\label{planetparams}
\begin{tabular}{lllll}
\hline
Parameter & Kepler-47 (Star B) & Kepler-47b & Kepler-47c \\
\hline
\hline
Semi-major axis (AU) & $0.0836 \pm 0.0014$ & $0.2956 \pm 0.0047$ & $0.989 \pm 0.016$ \\
Eccentricity & $0.0234 \pm 0.001$ & $0.034$ & $0.41$ \\
Inclination (deg.) & $89.34 \pm 0.12$ & $89.59 \pm 0.50$ & $89.826 \pm 0.010$  \\
Argument of pericenter (deg.) & $212.3 \pm 4.4$ & $0.0$ (fixed) & $0.0$ (fixed) \\
Longitude of node (deg.) & $0.0$ (fixed) & $0.0$ (fixed) & $0.0$ (fixed)  \\
Mean anomaly (deg.) & $0.0$ (fixed) &  $0.0$ (fixed) & $0.0$ (fixed)  \\
Orbital period (days) & $\simeq 7.5$ & $\simeq 49.5$ & $\simeq 303.2$  \\
Mass & $0.362~M_{\odot}$ & $10~M_{\bigoplus}$ & $23~M_{\bigoplus}$ \\ 
Radius & $0.3506~R_{\odot} \pm 0.0063$ & $2.98~R_{\bigoplus} \pm 0.12$ & $4.61~R_{\bigoplus} \pm 0.20$ \\
\hline
\end{tabular}
\end{table*}

\clearpage

\begin{table}
\centering
\caption{Mean-motion resonances between Kepler-47c and the third planet (d) when this object is in a circular orbit 
between planets b and c.} 
\tablecomments{Quantities $a_{\textnormal{inner}}$ and $a_{\textnormal{outer}}$ denote the inner and the outer 
boundary of the quasi-periodic regions shown in figure ~\ref{K47_Map055_Map056} for a circular orbit.}
\label{detailsofMMRinterior}
\begin{tabular}{lcc}
\hline
Resonance & $a_{\textnormal{inner}}$ [AU[ & $a_{\textnormal{outer}}$ [AU]\\
\hline
\hline
4c:7d & 0.677 & 0.680 \\
3c:5d & 0.700 & 0.704 \\
5c:8d & 0.720 & 0.723 \\
2c:3d & 0.749 & 0.757 \\
5c:7d & 0.786 & 0.791 \\
3c:4d & 0.811 & 0.818 \\
7c:9d & 0.833 & 0.835 \\
4c:5d & 0.847 & 0.853 \\
5c:6d & 0.871 & 0.875 \\
6c:7d & 0.888 & 0.893 \\
\hline
\end{tabular}
\end{table}

\vskip  1in

\begin{table}
\centering
\caption{Same as Table ~\ref{detailsofMMRinterior} with the exception that the third planet is in an orbit 
exterior to Kepler-47c.}
\tablecomments{Mean-motion resonances with semimajor axes larger than 1.6 AU have not been included.}
\label{detailsofMMRexterior}
\begin{tabular}{lcc}
\hline
Resonance & $a_{\textnormal{inner}}$ [AU] & $a_{\textnormal{outer}}$ [AU]\\
\hline
\hline
3d:4c & 1.189 & 1.199 \\
2d:3c & 1.279 & 1.303 \\
5d:8c & 1.346 & 1.352 \\
3d:5c & 1.385 & 1.386 \\
4d:7c & 1.424 & 1.437 \\
5d:9c & 1.471 & 1.478 \\
1d:2c & 1.538 & 1.595 \\
\hline
\end{tabular}
\end{table}

\end{document}